\documentclass[longauth]{aa}
\usepackage{euclid}
\usepackage{graphicx}
\usepackage{natbib}
\usepackage{scalerel}
\usepackage[table]{xcolor}
\usepackage{soul}
\usepackage{txfonts}
\usepackage[pdfencoding=auto,psdextra]{hyperref}
\hypersetup{
    colorlinks=true,
    linkcolor=blue,
    filecolor=magenta,      
    urlcolor=blue,
    citecolor=blue
}
\makeatletter
\renewcommand*\aa@pageof{, page \thepage{} of \pageref*{LastPage}}
\makeatother
\usepackage[utf8]{inputenc}
\usepackage[switch, modulo]{lineno}
\begin{document}
%
%
\title{\Euclid preparation. XXXII. Evaluating the weak lensing cluster mass biases using the Three Hundred Project hydrodynamical simulations}
\newcommand{\orcid}[1]{} 
\author{Euclid Collaboration: C.~Giocoli$^{1,2}$\thanks{\email{carlo.giocoli@inaf.it}}, M.~Meneghetti\orcid{0000-0003-1225-7084}$^{1,2}$, E.~Rasia$^{3,4}$, S.~Borgani\orcid{0000-0001-6151-6439}$^{3,5,6,4}$, G.~Despali\orcid{0000-0001-6150-4112}$^{7}$, G.~F.~Lesci\orcid{0000-0002-4607-2830}$^{8,1}$, F.~Marulli\orcid{0000-0002-8850-0303}$^{8,1,2}$, L.~Moscardini\orcid{0000-0002-3473-6716}$^{8,1,2}$, M.~Sereno\orcid{0000-0003-0302-0325}$^{1,2}$, W.~Cui\orcid{0000-0002-2113-4863}$^{9,10,11}$
% ...
}

\institute{$^{1}$ INAF-Osservatorio di Astrofisica e Scienza dello Spazio di Bologna, Via Piero Gobetti 93/3, 40129 Bologna, Italy\\
$^{2}$ INFN-Sezione di Bologna, Viale Berti Pichat 6/2, 40127 Bologna, Italy\\
$^{3}$ INAF-Osservatorio Astronomico di Trieste, Via G. B. Tiepolo 11, 34143 Trieste, Italy\\
$^{4}$ IFPU, Institute for Fundamental Physics of the Universe, via Beirut 2, 34151 Trieste, Italy\\
$^{5}$ Dipartimento di Fisica - Sezione di Astronomia, Universit\'a di Trieste, Via Tiepolo 11, 34131 Trieste, Italy\\
$^{6}$ INFN, Sezione di Trieste, Via Valerio 2, 34127 Trieste TS, Italy\\
$^{7}$ Institut f\"{u}r Theoretische Astrophysik, Zentrum f\"{u}r Astronomie, Heidelberg Universit\"{a}t, Albert-Ueberle-Str. 2, 69120, Heidelberg, Germany\\
$^{8}$ Dipartimento di Fisica e Astronomia "Augusto Righi" - Alma Mater Studiorum Universit\`{a} di Bologna, via Piero Gobetti 93/2, 40129 Bologna, Italy\\
$^{9}$ Departamento de F\'isica Te\'orica, Facultad de Ciencias, Universidad Aut\'onoma de Madrid, 28049 Cantoblanco, Madrid, Spain\\
$^{10}$ Centro de Investigaci\'{o}n Avanzada en F\'isica Fundamental (CIAFF), Facultad de Ciencias, Universidad Aut\'{o}noma de Madrid, 28049 Madrid, Spain\\
$^{11}$ Institute for Astronomy, University of Edinburgh, Royal Observatory, Blackford Hill, Edinburgh EH9 3HJ, UK\\
$^{12}$ International Centre for Radio Astronomy Research, University of Western Australia, 35 Stirling Highway, Crawley, Western Australia 6009, Australia\\
$^{13}$ Laboratoire Univers et Th\'{e}orie, Observatoire de Paris, Universit\'{e} PSL, Universit\'{e} Paris Cit\'{e}, CNRS, 92190 Meudon, France\\
$^{14}$ AIM, CEA, CNRS, Universit\'{e} Paris-Saclay, Universit\'{e} de Paris, 91191 Gif-sur-Yvette, France\\
$^{15}$ Institut f\"ur Astro- und Teilchenphysik, Universit\"at Innsbruck, Technikerstr. 25/8, 6020 Innsbruck, Austria\\
$^{16}$ Argelander-Institut f\"ur Astronomie, Universit\"at Bonn, Auf dem H\"ugel 71, 53121 Bonn, Germany\\
$^{17}$ Universit\'e Paris-Saclay, Universit\'e Paris Cit\'e, CEA, CNRS, AIM, 91191, Gif-sur-Yvette, France\\
$^{18}$ Universit\'e Paris-Saclay, CNRS, Institut d'astrophysique spatiale, 91405, Orsay, France\\
$^{19}$ Institut f\"ur Theoretische Physik, University of Heidelberg, Philosophenweg 16, 69120 Heidelberg, Germany\\
$^{20}$ Max Planck Institute for Extraterrestrial Physics, Giessenbachstr. 1, 85748 Garching, Germany\\
$^{21}$ INAF-Osservatorio Astrofisico di Torino, Via Osservatorio 20, 10025 Pino Torinese (TO), Italy\\
$^{22}$ Dipartimento di Fisica, Universit\`{a} di Genova, Via Dodecaneso 33, 16146, Genova, Italy\\
$^{23}$ INFN-Sezione di Roma Tre, Via della Vasca Navale 84, 00146, Roma, Italy\\
$^{24}$ Department of Physics "E. Pancini", University Federico II, Via Cinthia 6, 80126, Napoli, Italy\\
$^{25}$ Instituto de Astrof\'isica e Ci\^encias do Espa\c{c}o, Universidade do Porto, CAUP, Rua das Estrelas, PT4150-762 Porto, Portugal\\
$^{26}$ Dipartimento di Fisica, Universit\'a degli Studi di Torino, Via P. Giuria 1, 10125 Torino, Italy\\
$^{27}$ INFN-Sezione di Torino, Via P. Giuria 1, 10125 Torino, Italy\\
$^{28}$ INAF-IASF Milano, Via Alfonso Corti 12, 20133 Milano, Italy\\
$^{29}$ Institut de F\'{i}sica d'Altes Energies (IFAE), The Barcelona Institute of Science and Technology, Campus UAB, 08193 Bellaterra (Barcelona), Spain\\
$^{30}$ Port d'Informaci\'{o} Cient\'{i}fica, Campus UAB, C. Albareda s/n, 08193 Bellaterra (Barcelona), Spain\\
$^{31}$ Institut d'Estudis Espacials de Catalunya (IEEC), Carrer Gran Capit\'a 2-4, 08034 Barcelona, Spain\\
$^{32}$ Institute of Space Sciences (ICE, CSIC), Campus UAB, Carrer de Can Magrans, s/n, 08193 Barcelona, Spain\\
$^{33}$ INAF-Osservatorio Astronomico di Roma, Via Frascati 33, 00078 Monteporzio Catone, Italy\\
$^{34}$ INAF-Osservatorio Astronomico di Capodimonte, Via Moiariello 16, 80131 Napoli, Italy\\
$^{35}$ INFN section of Naples, Via Cinthia 6, 80126, Napoli, Italy\\
$^{36}$ Centre National d'Etudes Spatiales, Toulouse, France\\
$^{37}$ Institut national de physique nucl\'eaire et de physique des particules, 3 rue Michel-Ange, 75794 Paris C\'edex 16, France\\
$^{38}$ Jodrell Bank Centre for Astrophysics, Department of Physics and Astronomy, University of Manchester, Oxford Road, Manchester M13 9PL, UK\\
$^{39}$ ESAC/ESA, Camino Bajo del Castillo, s/n., Urb. Villafranca del Castillo, 28692 Villanueva de la Ca\~nada, Madrid, Spain\\
$^{40}$ European Space Agency/ESRIN, Largo Galileo Galilei 1, 00044 Frascati, Roma, Italy\\
$^{41}$ Univ Lyon, Univ Claude Bernard Lyon 1, CNRS/IN2P3, IP2I Lyon, UMR 5822, 69622, Villeurbanne, France\\
$^{42}$ Institute of Physics, Laboratory of Astrophysics, Ecole Polytechnique F\'{e}d\'{e}rale de Lausanne (EPFL), Observatoire de Sauverny, 1290 Versoix, Switzerland\\
$^{43}$ Mullard Space Science Laboratory, University College London, Holmbury St Mary, Dorking, Surrey RH5 6NT, UK\\
$^{44}$ Departamento de F\'isica, Faculdade de Ci\^encias, Universidade de Lisboa, Edif\'icio C8, Campo Grande, PT1749-016 Lisboa, Portugal\\
$^{45}$ Instituto de Astrof\'isica e Ci\^encias do Espa\c{c}o, Faculdade de Ci\^encias, Universidade de Lisboa, Campo Grande, 1749-016 Lisboa, Portugal\\
$^{46}$ Department of Astronomy, University of Geneva, ch. d'Ecogia 16, 1290 Versoix, Switzerland\\
$^{47}$ INFN-Padova, Via Marzolo 8, 35131 Padova, Italy\\
$^{48}$ Universit\'e Paris-Saclay, Universit\'e Paris Cit\'e, CEA, CNRS, Astrophysique, Instrumentation et Mod\'elisation Paris-Saclay, 91191 Gif-sur-Yvette, France\\
$^{49}$ INAF-Osservatorio Astronomico di Padova, Via dell'Osservatorio 5, 35122 Padova, Italy\\
$^{50}$ Universit\"ats-Sternwarte M\"unchen, Fakult\"at f\"ur Physik, Ludwig-Maximilians-Universit\"at M\"unchen, Scheinerstrasse 1, 81679 M\"unchen, Germany\\
$^{51}$ Institute of Theoretical Astrophysics, University of Oslo, P.O. Box 1029 Blindern, 0315 Oslo, Norway\\
$^{52}$ Jet Propulsion Laboratory, California Institute of Technology, 4800 Oak Grove Drive, Pasadena, CA, 91109, USA\\
$^{53}$ Technical University of Denmark, Elektrovej 327, 2800 Kgs. Lyngby, Denmark\\
$^{54}$ Cosmic Dawn Center (DAWN), Denmark\\
$^{55}$ Max-Planck-Institut f\"ur Astronomie, K\"onigstuhl 17, 69117 Heidelberg, Germany\\
$^{56}$ Aix-Marseille Universit\'e, CNRS/IN2P3, CPPM, Marseille, France\\
$^{57}$ Universit\'e de Gen\`eve, D\'epartement de Physique Th\'eorique and Centre for Astroparticle Physics, 24 quai Ernest-Ansermet, CH-1211 Gen\`eve 4, Switzerland\\
$^{58}$ Department of Physics, P.O. Box 64, 00014 University of Helsinki, Finland\\
$^{59}$ Helsinki Institute of Physics, Gustaf H{\"a}llstr{\"o}min katu 2, University of Helsinki, Helsinki, Finland\\
$^{60}$ NOVA optical infrared instrumentation group at ASTRON, Oude Hoogeveensedijk 4, 7991PD, Dwingeloo, The Netherlands\\
$^{61}$ Department of Physics, Institute for Computational Cosmology, Durham University, South Road, DH1 3LE, UK\\
$^{62}$ Universit\'e C\^{o}te d'Azur, Observatoire de la C\^{o}te d'Azur, CNRS, Laboratoire Lagrange, Bd de l'Observatoire, CS 34229, 06304 Nice cedex 4, France\\
$^{63}$  Universit\'e Paris Cit\'e, CNRS, Astroparticule et Cosmologie, 75013 Paris, France\\
$^{64}$ European Space Agency/ESTEC, Keplerlaan 1, 2201 AZ Noordwijk, The Netherlands\\
$^{65}$ Leiden Observatory, Leiden University, Niels Bohrweg 2, 2333 CA Leiden, The Netherlands\\
$^{66}$ Kapteyn Astronomical Institute, University of Groningen, PO Box 800, 9700 AV Groningen, The Netherlands\\
$^{67}$ Department of Physics and Astronomy, University of Aarhus, Ny Munkegade 120, DK-8000 Aarhus C, Denmark\\
$^{68}$ Space Science Data Center, Italian Space Agency, via del Politecnico snc, 00133 Roma, Italy\\
$^{69}$ Institute of Space Science, Bucharest, 077125, Romania\\
$^{70}$ Dipartimento di Fisica e Astronomia "G.Galilei", Universit\'a di Padova, Via Marzolo 8, 35131 Padova, Italy\\
$^{71}$ Dipartimento di Fisica e Astronomia, Universit\'a di Bologna, Via Gobetti 93/2, 40129 Bologna, Italy\\
$^{72}$ Departamento de F\'isica, FCFM, Universidad de Chile, Blanco Encalada 2008, Santiago, Chile\\
$^{73}$ Institut de Ciencies de l'Espai (IEEC-CSIC), Campus UAB, Carrer de Can Magrans, s/n Cerdanyola del Vall\'es, 08193 Barcelona, Spain\\
$^{74}$ Centro de Investigaciones Energ\'eticas, Medioambientales y Tecnol\'ogicas (CIEMAT), Avenida Complutense 40, 28040 Madrid, Spain\\
$^{75}$ Instituto de Astrof\'isica e Ci\^encias do Espa\c{c}o, Faculdade de Ci\^encias, Universidade de Lisboa, Tapada da Ajuda, 1349-018 Lisboa, Portugal\\
$^{76}$ Universidad Polit\'ecnica de Cartagena, Departamento de Electr\'onica y Tecnolog\'ia de Computadoras, 30202 Cartagena, Spain\\
$^{77}$ Institut de Recherche en Astrophysique et Plan\'etologie (IRAP), Universit\'e de Toulouse, CNRS, UPS, CNES, 14 Av. Edouard Belin, 31400 Toulouse, France\\
$^{78}$ Infrared Processing and Analysis Center, California Institute of Technology, Pasadena, CA 91125, USA\\
$^{79}$ INAF-Osservatorio Astronomico di Brera, Via Brera 28, 20122 Milano, Italy\\
$^{80}$ Instituto de Astrof\'isica de Canarias, Calle V\'ia L\'actea s/n, 38204, San Crist\'obal de La Laguna, Tenerife, Spain\\
$^{81}$ Center for Computational Astrophysics, Flatiron Institute, 162 5th Avenue, 10010, New York, NY, USA\\
$^{82}$ School of Physics and Astronomy, Cardiff University, The Parade, Cardiff, CF24 3AA, UK\\
$^{83}$ INAF-Istituto di Astrofisica e Planetologia Spaziali, via del Fosso del Cavaliere, 100, 00100 Roma, Italy\\
$^{84}$ Ernst-Reuter-Str. 4e, 31224 Peine, Germany\\
$^{85}$ Department of Physics and Helsinki Institute of Physics, Gustaf H\"allstr\"omin katu 2, 00014 University of Helsinki, Finland\\
$^{86}$ Dipartimento di Fisica e Astronomia "Augusto Righi" - Alma Mater Studiorum Universit\'a di Bologna, Viale Berti Pichat 6/2, 40127 Bologna, Italy\\
$^{87}$ Institut d'Astrophysique de Paris, 98bis Boulevard Arago, 75014, Paris, France\\
$^{88}$ Junia, EPA department, 59000 Lille, France\\
$^{89}$ INFN-Bologna, Via Irnerio 46, 40126 Bologna, Italy\\
$^{90}$ Instituto de F\'isica Te\'orica UAM-CSIC, Campus de Cantoblanco, 28049 Madrid, Spain\\
$^{91}$ CERCA/ISO, Department of Physics, Case Western Reserve University, 10900 Euclid Avenue, Cleveland, OH 44106, USA\\
$^{92}$ Laboratoire de Physique de l'\'Ecole Normale Sup\'erieure, ENS, Universit\'e PSL, CNRS, Sorbonne Universit\'e, 75005 Paris, France\\
$^{93}$ Observatoire de Paris, Universit\'e PSL, Sorbonne Universit\'e, LERMA, 750 Paris, France\\
$^{94}$ Astrophysics Group, Blackett Laboratory, Imperial College London, London SW7 2AZ, UK\\
$^{95}$ SISSA, International School for Advanced Studies, Via Bonomea 265, 34136 Trieste TS, Italy\\
$^{96}$ Dipartimento di Fisica e Scienze della Terra, Universit\'a degli Studi di Ferrara, Via Giuseppe Saragat 1, 44122 Ferrara, Italy\\
$^{97}$ Istituto Nazionale di Fisica Nucleare, Sezione di Ferrara, Via Giuseppe Saragat 1, 44122 Ferrara, Italy\\
$^{98}$ Institut de Physique Th\'eorique, CEA, CNRS, Universit\'e Paris-Saclay 91191 Gif-sur-Yvette Cedex, France\\
$^{99}$ Institut d'Astrophysique de Paris, UMR 7095, CNRS, and Sorbonne Universit\'e, 98 bis boulevard Arago, 75014 Paris, France\\
$^{100}$ NASA Ames Research Center, Moffett Field, CA 94035, USA\\
$^{101}$ INAF, Istituto di Radioastronomia, Via Piero Gobetti 101, 40129 Bologna, Italy\\
$^{102}$ Institute for Theoretical Particle Physics and Cosmology (TTK), RWTH Aachen University, 52056 Aachen, Germany\\
$^{103}$ Institute for Astronomy, University of Hawaii, 2680 Woodlawn Drive, Honolulu, HI 96822, USA\\
$^{104}$ Department of Physics \& Astronomy, University of California Irvine, Irvine CA 92697, USA\\
$^{105}$ University of Lyon, UCB Lyon 1, CNRS/IN2P3, IUF, IP2I Lyon, France\\
$^{106}$ INFN-Sezione di Genova, Via Dodecaneso 33, 16146, Genova, Italy\\
$^{107}$ Aix-Marseille Universit\'e, CNRS, CNES, LAM, Marseille, France\\
$^{108}$ Department of Astronomy \& Physics and Institute for Computational Astrophysics, Saint Mary's University, 923 Robie Street, Halifax, Nova Scotia, B3H 3C3, Canada\\
$^{109}$ University Observatory, Faculty of Physics, Ludwig-Maximilians-Universit{\"a}t, Scheinerstr. 1, 81679 Munich, Germany\\
$^{110}$ Ruhr University Bochum, Faculty of Physics and Astronomy, Astronomical Institute (AIRUB), German Centre for Cosmological Lensing (GCCL), 44780 Bochum, Germany\\
$^{111}$ Department of Physics, Lancaster University, Lancaster, LA1 4YB, UK\\
$^{112}$ Univ. Grenoble Alpes, CNRS, Grenoble INP, LPSC-IN2P3, 53, Avenue des Martyrs, 38000, Grenoble, France\\
$^{113}$ Department of Physics and Astronomy, University College London, Gower Street, London WC1E 6BT, UK\\
$^{114}$ Department of Physics and Astronomy, Vesilinnantie 5, 20014 University of Turku, Finland\\
$^{115}$ Centre de Calcul de l'IN2P3, 21 avenue Pierre de Coubertin 69627 Villeurbanne Cedex, France\\
$^{116}$ Dipartimento di Fisica, Sapienza Universit\`a di Roma, Piazzale Aldo Moro 2, 00185 Roma, Italy\\
$^{117}$ University of Applied Sciences and Arts of Northwestern Switzerland, School of Engineering, 5210 Windisch, Switzerland\\
$^{118}$ INFN-Sezione di Roma, Piazzale Aldo Moro, 2 - c/o Dipartimento di Fisica, Edificio G. Marconi, 00185 Roma, Italy\\
$^{119}$ Centro de Astrof\'{\i}sica da Universidade do Porto, Rua das Estrelas, 4150-762 Porto, Portugal\\
$^{120}$ Institute of Cosmology and Gravitation, University of Portsmouth, Portsmouth PO1 3FX, UK\\
$^{121}$ Department of Mathematics and Physics E. De Giorgi, University of Salento, Via per Arnesano, CP-I93, 73100, Lecce, Italy\\
$^{122}$ INFN, Sezione di Lecce, Via per Arnesano, CP-193, 73100, Lecce, Italy\\
$^{123}$ INAF-Sezione di Lecce, c/o Dipartimento Matematica e Fisica, Via per Arnesano, 73100, Lecce, Italy\\
$^{124}$ CEA Saclay, DFR/IRFU, Service d'Astrophysique, Bat. 709, 91191 Gif-sur-Yvette, France\\
$^{125}$ Institute for Computational Science, University of Zurich, Winterthurerstrasse 190, 8057 Zurich, Switzerland\\
$^{126}$ Universit\'e St Joseph; Faculty of Sciences, Beirut, Lebanon\\
$^{127}$ Dipartimento di Fisica "Aldo Pontremoli", Universit\'a degli Studi di Milano, Via Celoria 16, 20133 Milano, Italy}
\abstract{ The photometric catalogue of galaxy clusters extracted from
  ESA \Euclid data is expected to be very competitive for cosmological
  studies.  Using state-of-the-art hydrodynamical simulations, we
  present systematic analyses simulating the expected weak lensing
  profiles from clusters in a variety of dynamic states and at
  wide range of redshifts.  In order to derive cluster masses, we use a
  model consistent with the implementation within the Euclid Consortium of the dedicated processing function and find that, when
  jointly modelling mass and the concentration parameter of the
  Navarro--Frenk--White halo profile, the weak lensing masses tend to be, on
  average,  biased low by 5--10\% with respect to the true mass, up to \mbox{$z=0.5$}.  Using a
  fixed value for the concentration $c_{200}=3$, the mass bias is
  diminished below 5\%, up to \mbox{$z=0.7$}, along with its relative uncertainty.  Simulating the weak
  lensing signal by projecting along the directions of the axes of the
  moment of inertia tensor ellipsoid, we find that orientation
  matters: when clusters are oriented along the major axis, the lensing signal is boosted, and the recovered weak lensing mass is
  correspondingly overestimated.  Typically, the weak lensing mass bias of individual clusters is modulated by the weak lensing
  signal-to-noise ratio, related to the redshift evolution of the number of galaxies 
  used for weak lensing measurements:  the negative mass bias tends to be larger
  toward higher redshifts.  However, when we use a fixed value of the
  concentration parameter, the redshift evolution trend is reduced.  These results
  provide a solid basis for the weak-lensing mass calibration required
  by the cosmological application of future cluster surveys from
  \Euclid and \emph{Rubin}.}  
\titlerunning{Simulated Weak Lensing Cluster Mass Biases}
\authorrunning{Giocoli et al. 2023}    
 \maketitle
\section{Introduction}
The abundance and the spatial distribution of galaxy clusters as a function of redshift represent important cosmological probes for
future wide-field surveys, and particularly for the ESA \Euclid
mission \citep{euclidredbook,scaramella22}.  Cluster cosmological
studies \citep[e.g.,][]{allen11} will complement the two main probes
based on weak lensing cosmic shear and galaxy clustering, improving
the figure of merit of the derived cosmological parameters \citep[see
  e.g.,][]{sartoris16}.

Galaxy clusters form close to the mass density peaks of the initial
matter density fluctuations and accrete mass through cosmic time as a
consequence of repeated merging events
\citep{tormen98a,tormen04,giocoli12b,kravborg12}.  In the present day,
clusters represent the largest virialised structures in the Universe.
Their structural properties \citep{giocoli08b,despali14,despali17} are
important tracers of their assembly history and dynamical state.  For
example, a lower value of their concentration parameter and a higher
mass fraction in substructures are typical of late-forming systems.
On the contrary, fewer substructures and high concentrations are
ordinary for clusters that assemble most of their mass at higher
redshifts, which appear more regular and are rounder
\citep{gao04,delucia04,giocoli10a,bonamigo15,mostoghiu19}.

The ESA \Euclid mission, complemented by the multiband photometric
support from the ground by various observational facilities will be
able to identify galaxy clusters using two complementary algorithms:
\texttt{AMICO} and \texttt{PZWav} \citep{adam19}.  While the
\texttt{AMICO} (Adaptive Matched Identifier of Clustered Objects)
algorithm \citep{bellagamba11,bellagamba18,maturi19} uses an enhanced
matched filter method that looks for cluster candidates by convolving
the 3D galaxy distribution with a redshift-dependent filter,
\texttt{PZWav} \citep{gonzalez14} is a wavelet transform-based code
that searches for overdensities on fixed physical scales.  The two
methods have been extensively studied thanks to dedicated activities
and performance challenges \citep{adam19} that guarantee both methods
have a purity and completeness of at least $80$ per cent for systems
with a mass larger than $M_{200}=10^{14}\,M_{\odot}$\footnote{
The mass within the radius that encloses 
200 times the critical density of the universe at a given redshift.} 
and redshift $z<2$.
Being complementary, the matching procedures of the two detection
algorithms are expected to generate a highly pure and complete
catalogue.

The use of clusters as a cosmological tool relies on the accuracy with
which we can recover their true mass \citep[][for a review]{pratt19}.
The mass enclosed within a given overdensity can be measured in
numerical simulations \citep{sheth99b,springel01b,tormen04}, and it is used to build up analytical functions to describe the number density
of haloes as a function of mass and redshift \citep{tinker08,
  despali16,klypin16,chua17,bocquet20,ondaromallea22,castro22}.  However, from an
observational perspective, we have to rely on mass proxy estimates
that could be scattered and biased. Typically, these biases and scatters are due to the
simple analytical models used to characterise the complexity of the
astrophysical processes taking place in the cluster environments
\citep{becker11,grandis21}. In addition, the scatter in the cluster
richness \footnote{Defined as the total number of member galaxies 
within the cluster radius $R_{200}$, above a typical characteristic luminosity.}-mass relation could increase because of different feedback
and quenching mechanisms taking place in the highly-dense cluster
environments. 

The dynamics of cluster galaxies has been extensively used in
multiple dedicated observations to characterise the cluster mass
\citep{biviano06,biviano13,bocquet15b,capasso19}.  This method may
suffer from the presence of interlopers that may systematically bias
the derived dynamical mass. This limitation could be alleviated by
removing clusters with significant evidence of sub-structures and
groups along the line of sight or by selecting early-type galaxies as
spectroscopic targets \citep{damsted23}.

The hydrostatic masses derived from X-ray observations and assuming a
model-based hydrostatic equilibrium are biased low with respect to the
true mass by around 20-25 per cent, as discussed in different studies
based on hydrodynamical simulations
\citep{lau09,meneghetti10b,rasia12,biffi16,ansarifard20,barnes21,gianfagna21}.
The inclusion of turbulent and non-thermal pressure support in the
model attenuates the discrepancies \citep{angelinelli20, ettori22},
but at the price of dedicated high-resolution multiwavelength
follow-up observations of the cluster environment.

Gravitational lensing, particularly in the weak regime, is the primary
method to be used within the Euclid Collaboration to weigh clusters.  
Since lensing does not rely on any assumption on the
dynamical state of the various mass components of the cluster, it
provides, in principle, an unbiased estimate of the projected matter
density distribution \citep{bartelmann01,bartelmann10}.  However, it
suffers from deprojection effects when converting the derived weak
lensing mass to the actual three-dimensional one
\citep{meneghetti08,giocoli12c,giocoli14}.  Dedicated works have been
recently carried out, based on state-of-the-art hydrodynamical
simulations, to assess the reliability of the weak lensing mass from
future wide-field surveys \citep{grandis19} and to quantify the
effects of baryonic physics \citep{henson17,grandis21,cormer22}. On the other hand, 
\citet{martizzi14,cusworth14,velliscig14,bocquet16a,debackere21,castro21} have studied  
the impact of baryons on the halo mass function and their effect on the cosmological 
parameter estimates using cluster counts. 
Typically, weak lensing mass biases are, in principle, not problematic, 
as long estimated from representative simulated cluster samples 
\citep[e.g.][]{applegate16,schrabback18,grandis19,sommer22}
and self-consistently accounted for in cluster scaling relation and cosmology analyses 
\citep[e.g.][]{dietrich19, bocquet19, schrabback21,chiu22,zohren22}.

Recent works by \citet{debackere22b,debackere22a} have shown the potential of weak lensing aperture mass to recover the projected mass
density distribution of galaxy cluster regions. The method does not
rely on any particular mass density profile model but needs to be tuned
and trained on large sample cosmological simulations.  This approach
also requires us to rewrite all likelihoods of the cluster cosmology
pipeline in terms of projected quantities instead of 3D masses
\citep[see also][]{giocoli12c}.

Thanks to the high efficiency, the unprecedented combination of
spatial resolution and sensitivity of the VIS instrument
\citep{cropper16}  onboard the \Euclid satellite, we are expected to
reach a number density of sources for weak lensing studies
of  30 galaxies per square arcminute.  This large number density of sources will give us the possibility to recover weak
lensing masses for individual massive clusters. However, going toward
high-redshift clusters $z>0.6$, the dearth of available sources
requires stacking their signals, for well-defined richness and
redshift bins, to increase the signal-to-noise and the accuracy of the
(stacked) mass.

In this paper, we use a large sample of cosmological hydrodynamic
simulations of clusters from The Three Hundred Collaboration
\citep{cui18}, to make a systematic assessment of the weak lensing
mass bias of clusters. We study how the modelling function parameters (truncation radius, concentration, etc.)
affect the recovered weak lensing mass; in addition, we investigate
how the mass bias depends on the true mass of the cluster, on its
redshift, and, particularly, on its orientation with respect to the
line of sight, which represents the largest source of intrinsic scattering at fixed cluster mass. 
We would like to underline that the results here presented 
have been obtained assuming a specific modelling of the feedback inside 
clusters, and they could be sensitive to 
this choice. However, \citet{grandis21} have estimated that different hydrodynamic
solvers may have an impact on the weak lensing mass bias below $5\%$. 
We have planned a future work to systematically assess this.

The paper is organised as follows: in Sect. \ref{300sims} we introduce
the cluster simulations summarising the cluster properties that we
will use for this work; in Sect. \ref{clwlsims} we present our lensing
pipeline and how we simulate the expected signal from the ESA \Euclid
data; Sect. \ref{anamodelres} discusses the results using our
modelling functions.  In Sect. \ref{summary} we summarise and
conclude.

\section{The Three Hundred simulation data-set}
\label{300sims}

In this work, we rely on dedicated weak lensing simulations of
clusters extracted from the re-simulated regions by the Three Hundred
Collaboration \citep{cui18,cui22}.  The data set consists of 324
regions centred on the most massive clusters ($M_{200} \gtrapprox 8
\times 10^{14}\,h^{-1}\,M_{\odot}$) identified at $z=0$ in the DM-only
MDPL2 MultiDark simulation \citep{klypin16}, with a box-size of $1\,$Gpc
on a side. The parent simulation was run adopting the cosmological
parameters as derived by the \textit{Planck} mission
\citep{planck16a}: $\Omega_{\rm m}=0.307$, $\Omega_{\rm b}=0.048$,
$\Omega_{\rm \Lambda}=0.693$, $h=0.678$, $\sigma_8=0.823$ and $n_{\rm
  s}=0.96$. For each selected region, initial conditions with multiple
levels of mass refinements were generated using the
\texttt{GINNUNGAGAP}
code\footnote{\url{https://github.com/ginnungagapgroup/ginnungagap}}.
Within the highest resolution Lagrangian region -- at least five times
larger than the cluster virial radius -- particles are divided into
dark matter (DM) and gas types according to the considered
cosmological baryon fraction: $m_{\rm dm}=12.7\times
10^8\,h^{-1}\,M_{\odot}$ and $m_{\rm gas}=2.36\times
10^8\,h^{-1}\,M_{\odot}$, respectively.

The resolution outside this region is degraded to reduce the
computational cost with respect to the parent original simulation. It
is worth mentioning that each re-simulated region, with a typical
radius of $15\,h^{-1}\,\mathrm{Mpc}$ from the centre, may contain
additional groups and filaments not physically associated and not
gravitationally bound to the virialised cluster but important for
total projected mass quantity studies.

\begin{figure*}
  \includegraphics[width=0.43\hsize]{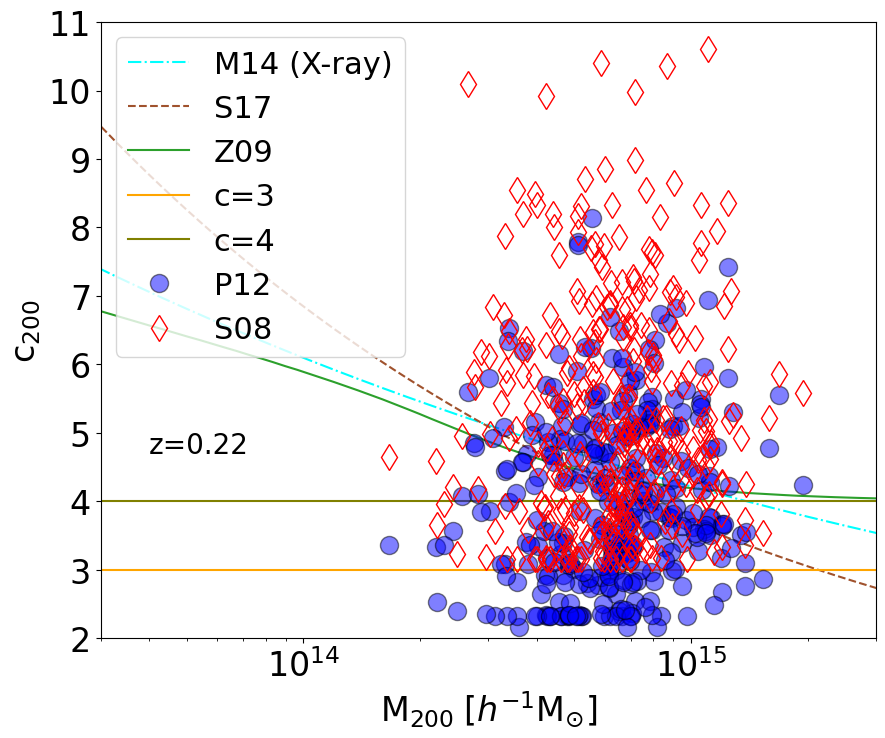}
  \includegraphics[width=0.57\hsize]{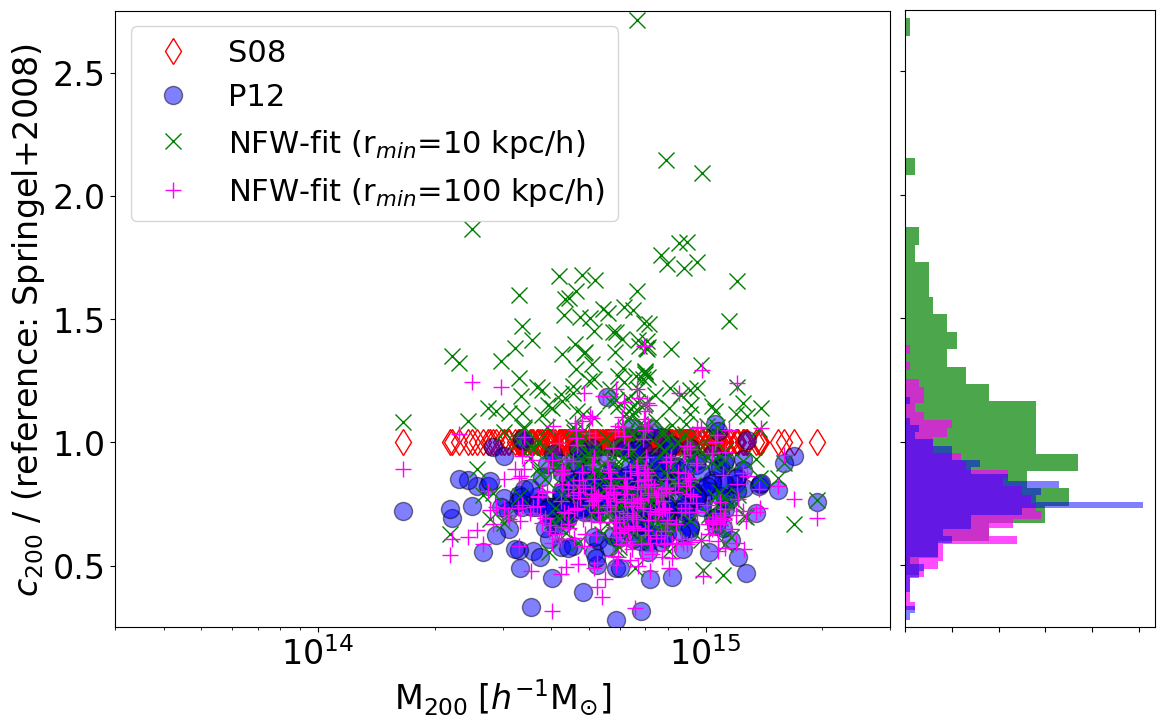}
  \caption{\textit{Left panel}: concentration-mass relation of the
    clusters at $z=0.22$. The blue and red data points represent the
    concentration values computed using the \citet[][hereafter
      P12]{prada12} and \citet[][hereafter S08]{springel08a}
    relations, respectively. The various lines display different
    concentration-mass relation models computed at
    $z=0.22$ \label{figconcentration}.  \textit{Right panel}: ratio
    between the measured concentration, using different methods
    compared with respect to the \citet{springel08a} formalism.  Green
    crosses and magenta pluses display the case in which we compute
    the concentration by fitting the differential logarithmic density
    profile outside 10, and $100\,h^{-1}\,\mathrm{kpc}$,
    respectively.}
\end{figure*}

The evolution of the particle distribution from the initial-conditions
($z=120$) until the present time was followed using \texttt{GADGET-X,}
based on the gravity solved \texttt{GADGET-3 }Tree-PM code. The code
uses an improved smooth-particle hydrodynamics (SPH) scheme
\citep{beck16b} to follow the evolution of the gas component with
artificial thermal diffusion, time-dependent artificial viscosity,
high-order Wendland C4 interpolating kernel and wake-up scheme. These
improvements increase the SPH capability of following gas-dynamical
instabilities and mixing processes by better describing the
discontinuities and reducing the clumpiness instability of gas.

\begin{figure*}
  \includegraphics[width=0.5\hsize]{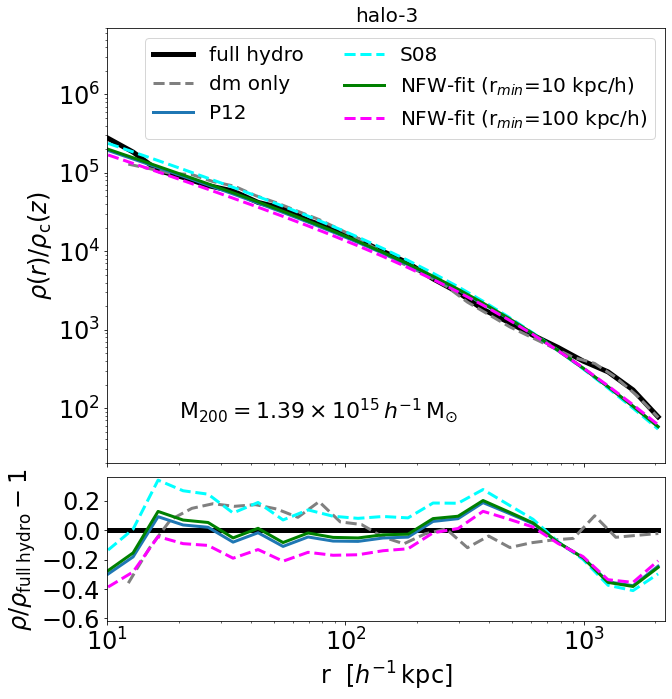}
  \includegraphics[width=0.5\hsize]{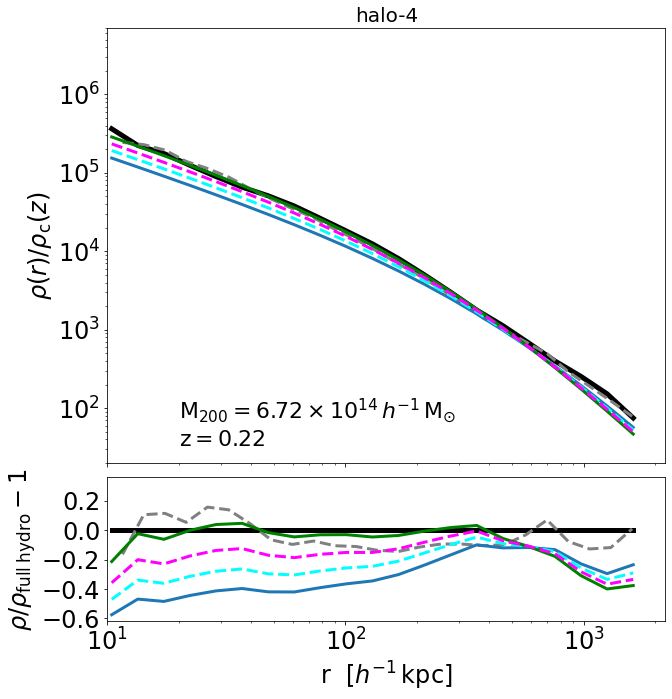}
  \caption{\label{fig3dprof}Spherical averaged density profiles of
    halo-3 and halo-4.  In each panel, the solid black curve displays
    the total matter density profile of the full hydrodynamical runs,
    as considered in this work. For comparison, the dashed grey curve
    shows the profile of the dark matter only runs. With solid blue and
    dashed cyan lines, we exhibit the NFW profiles assuming concentrations as
    computed using P12 and S08 formalisms, respectively. With solid
    green and dashed magenta lines, we show the NFW profiles where
    concentrations have been computed fitting the logarithm profile
    outside 10 and 100 $h^{-1}\,kpc$, respectively.  Profiles extend
    up to the halo virial radius $R_{200}$, as computed by
    \texttt{AHF}.}
\end{figure*}

As described in more detail in \citet{rasia15}, the simulations
include metallicity-dependent radiative cooling and the effect of a
uniform time-dependent UV background \citep{planelles14}.  The
sub-resolution star formation model follows \citet{springel03}, and
the metal production from SN-II, SN-Ia, and asymptotic-giant-branch
stars use the original recipe by \citet{tornatore07}. The active
galactic nucleus feedbacks and supermassive black hole accretions are
modelled with the implementations presented in \citet{steinborn15}.

\begin{figure*}[!h]
  \includegraphics[width=\hsize]{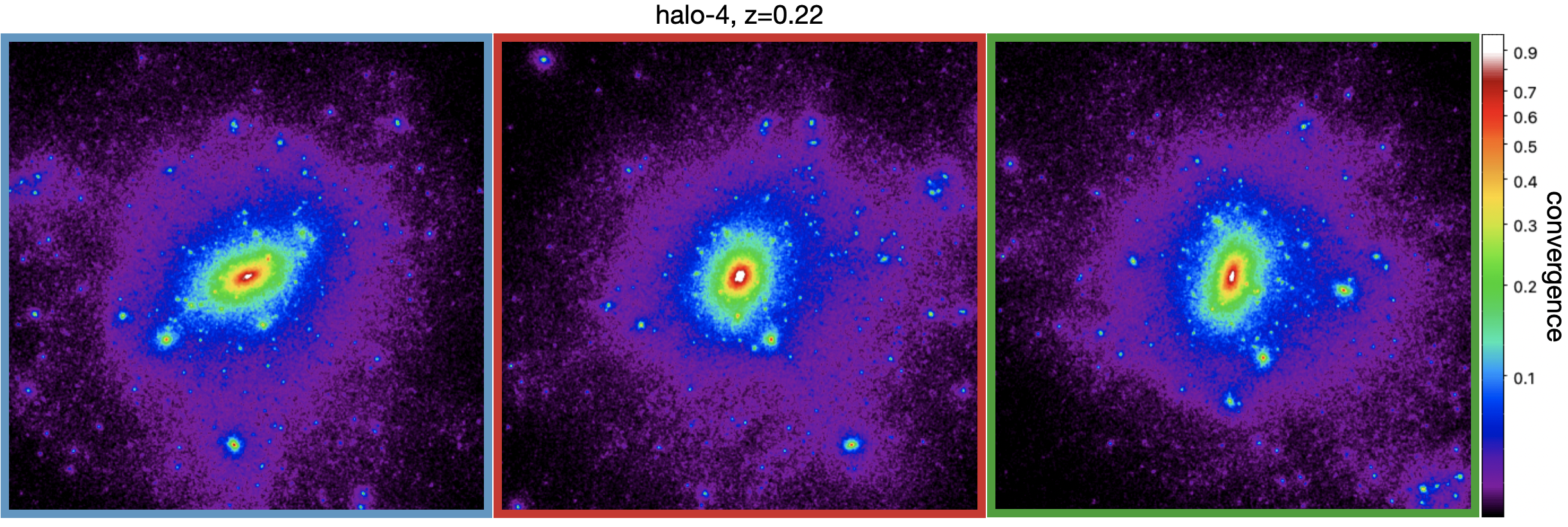}
  \caption{Convergence maps of the three considered random projections
    of a cluster, namely the halo-4 at $z=0.22$. The left, central and
    right panels show the projections along the $z$, $y$, and $x$ axis
    coordinates of the re-simulated box.  The source redshift of the
    map is fixed at $z_{\rm s}=3$.  The size of the field of view is
    $5\,\mathrm{Mpc}$ ($3.4\, h^{-1}\,\mathrm{Mpc}$) on a side,
    $10\,\mathrm{Mpc}$ along the line-of-sight.  The map resolution is 2048 pixels on a side.\label{figkappa}}
\end{figure*}

At each simulation snapshot, the haloes were identified using
\texttt{AHF} \citep[Amiga Halo Finder:][]{knollmann09} which
consistently accounts for DM, star, and gas particles in finding and
characterising halo properties. For each halo, the algorithm defines
$M_{200}$, using a spherical overdensity algorithm, i.e., the mass
within the radius $R_{200}$ which encloses 200 times the critical
comoving density of the universe $\rho_{\rm c}(z)$ at the
corresponding redshift,
\begin{equation}
M_{200} = \dfrac{4 \pi}{3} R_{200}^3\, 200 \; \rho_{\rm c}(z)\,.
\end{equation}
For each halo, we consider subhaloes whose centres lie within the
corresponding halo radius $R_{200}$.  The subhalo properties are
defined at the truncation radius $R_{\rm t}$ (the outer edge of the
system) and hence mass, density profile, velocity dispersion, rotation
curve are calculated using the gravitationally bound particles inside
this radius \citep{tormen98b}.  In particular, for each halo in this
work we make use of:
\begin{itemize}
\item the mass fraction in substructures $f_{\rm sub}$ accounting for
  the full subhalo hierarchy of subhaloes within subhaloes;
\item the halo virial circular velocity $V_{200} = \sqrt{\dfrac{G
    M_{200}}{R_{200}}}$;
\item the maximum circular velocity $V_{\rm max}$;
\item the radius $R_{\rm max}$ at which the maximum circular velocity
  is attained;
\item the centre-of-mass offset $x_{\rm off}$ -- expressed in the unit
  of the halo radius $R_{200}$, defined as the difference between the
  centre-of-mass and the maximum density peak of the halo, which we
  denote as the cluster centre;
\item the virial ratio $\eta \equiv \left(2 T - E_{\rm s} \right)
  /|W|$, where $T$ indicates the total kinetic energy, $E_{\rm s}$ the
  energy from surface pressure and $W$ refers to the total potential
  energy;
\item the eigenvalues and the eigenvectors of the moment of inertia
  tensor.
 \end{itemize}

The Navarro--Frenk--White (hereafter NFW) density profile
\citep{navarro96,navarro97}, depending on the host halo mass and
concentration is defined as:
\begin{equation}
  \rho_{\rm NFW}(r|M_{200},c_{200})=\frac{\rho_{\rm s}}{(r/r_{\rm
      s})(1+r/r_{\rm s})^2} \label{NFW}
\end{equation}
where $\rho_{\rm s}$ is defined as:
\begin{equation}
   \rho_{\rm s} = \dfrac{M_{200}}{4 \pi r_{\rm s}^3} \dfrac{1}{\ln(1+c_{200}) -
     c_{200}/(1+c_{200})}\;,
\end{equation}
and $r_{\rm s}=R_{200}/c_{200}$. 

From the global halo quantities, we adopt two prescriptions to define
the halo concentration.  Assuming that the halo density profile
follows an NFW relation, \citet[][hereafter S08]{springel08b} defines
the halo concentration numerically by solving
\begin{equation}
\dfrac{200}{3} \dfrac{c_{200}^3}{\ln(1+c_{200}) - c_{200}/(1+c_{200})} =
14.426 \left( \dfrac{V_{\rm max}}{H(z)\, R_{\rm max}} \right)^2\,,
\label{eqSpringel}
\end{equation}
where $H(z)$ represents the Hubble parameter at the corresponding
redshift.

Following a simpler version of the density profile parametrisation,
\citet[][hereafter P12]{prada12} describe the halo concentration in
terms of the velocity ratio only as follows:
\begin{equation}
  \dfrac{V_{\rm max}}{V_{200}} = \sqrt{\dfrac{0.216 \; c_{200}}
    {\ln(1+c_{200}) -c_{200}/(1+c_{200})} }\,.
\label{eqPrada}
\end{equation}
The two above definitions of concentration coincide if the halo is
perfectly defined by a spherical NFW profile.

Different post-processing analyses of numerical simulations 
have historically adopted different concentration definitions. In this paragraph, we compare
models based on different concentration definitions employed in our study.
The left panel of Fig.\ref{figconcentration} displays the
concentration-mass relation for the Three Hundred clusters at
$z=0.22$. The red diamonds and the blue-filled circles refer to the
concentration parameters computed using Eq.~\eqref{eqSpringel} and
\eqref{eqPrada}, respectively. The figure also contains various
predictions at the corresponding redshift, obtained after modelling the results from  
numerical simulations  \citep[][hereafter Z09, M14, respectively]{zhao09,meneghetti14}
or interpreting observational data \citep[][hereafter S17]{sereno17}. In particular, for
the Z09 model we adopt the \citet{giocoli12b} formalism to follow the
main halo mass accretion history back in time.  From the figure, we
notice that the two adopted models for the concentration predict a
different value at fixed halo mass: at this redshift on average
\citet{prada12} underestimates the concentration parameter by
approximately $20\%$ with respect to \citet{springel08b}. This is a
manifestation of the fact that the clusters, particularly in
hydrodynamical simulations, deviate from the perfect NFW profile for
which the two models would have the same concentration.  In addition,
the S08 model, written in terms of the radius $R_{\rm max}$, tends to
be more sensitive to baryonic physics and adiabatic contraction than
P12, which is parametrised only in terms of velocities, the average
relative difference between the two models varies with redshift.  In
the right panel of Fig. \ref{figconcentration}, we display the ratio
between various concentration definitions with respect to the
prediction by \citet{springel08b}.  Green crosses and magenta plus
signs correspond to the concentration computed by fitting the
logarithm of the total differential density profile, computed by the
\texttt{AHF}, outside 10 and 100 $h^{-1}$kpc from the centre,
respectively: binning procedures and reference models may impact the
concentration definition \citep{meneghetti13b}.  Density profiles are
very sensitive to baryonic effects at small radii. Higher
concentration parameters  result from a steepening of the inner slope
of the profile \citep{schaller15a,schaller15b,ahad21,jung22} due to
the adiabatic contraction, whose effect is mitigated by the AGN
feedback but not completely suppressed \citep{rasia13}.  Since this
paper aims at studying the WL mass reconstruction and the projected
density profiles are not reliable within the central 100 kpc --
noticing that the corresponding concentrations are in good agreement
with P12. For the rest of this work, we will also use S08 as a reference 
to give an idea of the theoretical uncertainties.

In Fig.~\ref{fig3dprof}, we show the spherical average density profiles
of two clusters: halo-3 (on the left) and halo-4 (on the right). In
each panel, the black solid and dashed grey curves refer to the total
density profile of the full hydrodynamical simulations and of the
corresponding dark matter-only runs. Profiles extend up to $R_{200}$ 
defined as the radius enclosing 200 times the critical density.  Solid blue and
dashed cyan curves display the NFW profiles using the concentration
values as derived from the P12 and S08 relations, respectively. With solid
green and dashed magenta lines, we display the NFW profiles where
concentration has been derived modelling the logarithmic density
profile outside 10 and 100 $h^{-1}\, kpc$, respectively.  From the
figure, we notice that while for halo-3 the solid blue and green lines
describe relatively well the total mass density profile of the cluster,
for halo-4 only the green line well represents the matter density
distribution. This could be due to the fact that while the halo-3 has
an inner slope $\mathrm{d} \log\rho/\mathrm{d} \log r$ close to -1,
the cluster halo-4 has a steeper slope toward the centre. We can also notice 
that the density in the outskirts of the clusters is higher than expected 
from the NFW profile due to the presence of substructures. Those 
differences reduce when looking at projected quantities, highlighting that
the  NFW prescription is sufficient for the analysis in this paper. 

We would also like to underline that the extrapolation of those 3D uncertainties into projected quantities 
 is not straightforward and may require some reference concentration-mass relation model to be adopted \citep{hoekstra13,simet17b,simet17,kiiveri21,sommer22}. We will describe how those impact the recovered weak lensing mass bias in the next section. 

\section{Cluster weak lensing simulations}
\label{clwlsims}

In order to build up the mass maps, we proceed as follows.  From the
\texttt{AHF} catalogues, we read the positions of the halo centres in
comoving units ($x_{\rm c},\,y_{\rm c},\,z_{\rm c}$) and then the
particle positions with associated masses and types from the
corresponding snapshot file. Each snapshot contains dark matter, gas,
star, and black hole particles. For each cluster, we consider three
random lines of sight corresponding to the axes of the simulation box,
along which we project the cluster particles on a perpendicular plane
centred on the cluster centre. We consider particles in a slice of
depth $\pm 5\, \mathrm{Mpc}$ ($3.4\,h^{-1}\,\mathrm{Mpc}$) in front
and behind the cluster.  The maps have a final size of
$5\,\mathrm{Mpc}$ on a side, resolved with $2048^2$ pixels and are
produced using \texttt{Py-SPHViewer} \citep[for more details we refer
  to][]{benitez15}. The choice of the size of the field of view is
motivated by the fact that we are mainly interested in modelling the
projected matter density distribution of the main cluster without
being much affected by the additional source of uncertainty associated
to the large-scale matter density distribution along the line of sight
\citep{hoekstra01,hoekstra03}. This value also represents a reasonable 
compromise to avoid influence due to low-resolution particles in the re-simulated box.  
In addition, it is worth stressing that the scatter in the weak lensing mass bias could be underestimated
with respect to what is measured by \citet{becker11}, who used a larger line-of-sight integration.

The convergence $\kappa$ is obtained from the mass map by dividing the
mass per pixel by its associated area to obtain the surface density
$\Sigma(\vec{\theta})$ and by the critical surface density
$\Sigma_{\rm crit}$ \citep{bartelmann01} that can be read as:
\begin{equation}
  \Sigma_{\rm crit} \equiv \dfrac{c^2}{4 \pi G} \dfrac{D_{\rm s}}{D_{\rm l}
    D_{\rm ls}}\,,
  \label{eq_scrit}
\end{equation}
where $D_{\rm l}$, $D_{\rm s}$, and $D_{\rm ls}$ are the
observer-lens, observer-source, and source-lens angular diameter
distances, respectively; $c$ represents the speed of light and $G$ the
universal gravitational constant. The critical surface density is a function of the redshift of the lens and of the source. Our reference
weak lensing maps have been created assuming a fixed source redshift
$z_{\rm s}=3$.

In Fig.~\ref{figkappa}, we show three convergence maps of the same
cluster, namely halo-4 at $z=0.22$. The left, central and right panels
display the system projected along the $z$, $y$, and $x$-axis of the
comoving re-simulated box, respectively.  Since there is no particular
selection with respect to the triaxial properties of the cluster
haloes, these projections can be seen as \textit{random} projections
and as such we will name them as proj0, proj1, proj2, respectively, in
the following.

For each cluster, redshift, and projection, we generate weak lensing
convergence maps, specifically, we consider 9 redshifts from $z=0.12$
to $z=0.98$ as listed in Table \ref{tab_zsnaps}.

\begin{table}[h]  
\begin{center}
\caption{List of the analysed simulation snapshots (first column) and the corresponding
  redshifts (second column). The third column specifies the average
  number of background galaxies used to construct the simulated weak
  lensing shear profile as expected by the \Euclid-ESA mission to lay
  beyond a cluster at a given redshift.  The fourth expresses the
  square root of the average number of background galaxies rescaled by the ratio of
  angular diameter distances, relevant for quantifying lensing detectability.  The last column shows the average
  cluster mass at the corresponding redshift, in the unit of $10^{14}\,h^{-1}\,M_{\odot}$.}
\label{tab_zsnaps}
\begin{tabular}{lcccc}
\hline
snap.  &  $z$ & $n_{\rm g}^{a}$ &$\sqrt{n_{\rm g}}\, D_{\rm ls}/D_{\rm s}^{b}$ & $\langle M_{200} \rangle ^{c}$\\
\hline
123   &     0.12   &   30 & 5.04& $7.85$    \\
119   &     0.22   &   29 & 4.62& $6.77$   \\
115   &     0.33   &   28  & 4.20& $5.77$  \\
113   &     0.39   &   27  & 3.95& $5.32$ \\
110   &     0.49   &   26  & 3.60& $4.66$  \\
107   &     0.59   &   24   &3.22& $3.99$ \\
104   &     0.70   &   22   &2.65& $3.38$ \\
101   &     0.82   &   19   &2.40& $2.83$ \\
098   &     0.94   &   17   &2.07& $2.38$ \\ 
\hline
\end{tabular}
\end{center}
\footnotesize{
  $^{a}\,[arcmin^{-2}]$\\
  $^{b}\,[arcmin^{-1}]$\\
  $^{c}\,[10^{14}\,h^{-1}\,M_{\odot}]$
  }
\end{table}

From the convergence $\kappa$, we define the lensing potential $\psi$,
using the two-dimensional Poisson equation as:
\begin{equation}
  \Delta_{\theta} \psi(\vec{\theta}) = 2 \, \kappa
  (\vec{\theta})\,,
\end{equation}
that we numerically solve in Fourier space using the Fast Fourier
Transform (FFT) method.  Since the FFT algorithm assumes periodic
conditions on the boundaries of the map, we zero-pad the exterior of
the convergence map by $0.5\,\mathrm{Mpc}$ before computing the
discrete Fourier transform and then remove the region from the final
result.

From the lensing potential $\psi$, we can then define the two
components $(\gamma_1,\gamma_2)$ of the pseudo-vector shear, as:
\begin{eqnarray}
  \gamma_1(\vec{\theta}) &=& \dfrac{1}{2} \left(\dfrac{\partial^2
    \psi(\vec{\theta})}{\partial x^2} - \dfrac{\partial^2
    \psi(\vec{\theta})}{\partial y^2}
  \right)\;,\\ \gamma_2(\vec{\theta}) &=& \dfrac{\partial^2
    \psi(\vec{\theta})}{\partial x \,\partial y} \,,
\end{eqnarray}
where $x$ and $y$ represent the two components of the vector
$\vec{\theta}$.

For each cluster, we use the shear maps to simulate an observed
differential surface mass density profile by randomly sampling the
field of view with a given number density of expected background
sources. 

In our analysis, we adopt a background density of sources
for weak lensing following the predicted distribution for the ESA
\Euclid wide-field survey normalised to the total value of 30 galaxies
per square arcmin \citep{euclidredbook}.  In the third column of
Table~\ref{tab_zsnaps} we report the average number of background
sources beyond each considered cluster redshift, whose corresponding
snapshot number is displayed in the first column.  

\begin{figure}
  \includegraphics[width=\hsize]{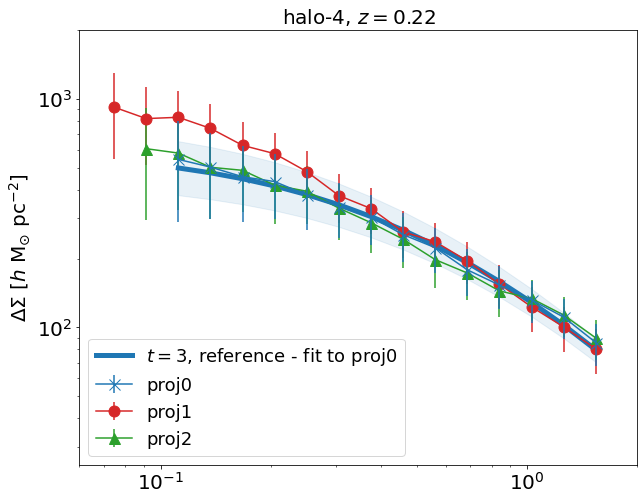}
  \includegraphics[width=\hsize]{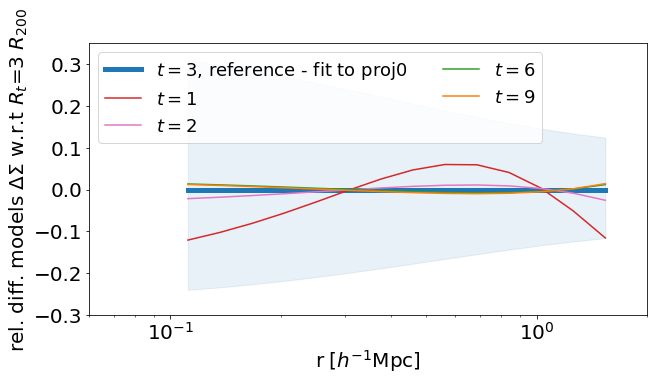}
  \caption{\textit{Top panel}: Excess surface mass density profiles of
    the three random projections for a cluster-size halo at $z_{\rm
      l}=0.22$, as displayed in Fig. \ref{figkappa}. The blue, red and
    green data points show the circular averaged profiles of the halo
    projected along the $z$, $y$, and $x$ axis of the re-simulation
    box. The error bars on the data points account for both the
    dispersion of the intrinsic ellipticity of background galaxies and
    the error associated with the average measure in the considered
    radial bin. The blue solid line displays the best-fit model to
    proj0, while the shaded region represents the model within the
    $68\%$ of confidence. \textit{Bottom panel}: relative difference
    of the best-fit models assuming different values of the truncation
    radius $R_{\rm t}$ with respect to the reference case with
    $t=3$.\label{figDsigma}}
\end{figure}

The approach of randomly sampling the shear field has been chosen to mimic 
the expected average profile of a cluster from a \Euclid wide-field exposure. Given 
the high number density of background sources, the profile has a much smaller scatter
with respect to the case in which we consider the complete shear field. We tested this for 
the halo-4 at $z=0.22$ and $z=0.94$, generating 10,000 random shear field realisations, finding 
results consistent with our reference case well within 1 $\sigma$ of the credibility regions of the posteriors.  

From the two components of the shear, we define the tangential shear
$\gamma_{\rm t}$ \citep{umetsu20b} at each sampled point $\theta_i$ of
the map, as:
\begin{equation}
\gamma_{\rm t} (\theta_i) =- \gamma_1(x_i,y_i) \cos(2 \phi_i) -
\gamma_2(x_i,y_i) \sin(2 \phi_i)\,,
\end{equation}
where $(0,0)$ is the centre of the cluster by construction, $\theta_i
= \smash{(x_i^2+y_i^2)^{1/2}}$ and $\phi_i = \arctan \left( y_i/x_i
\right)$.  This gives us the possibility to write the excess surface
mass density -- azimuthally averaging the measured quantities -- as:
\begin{equation}
\Delta \Sigma (\theta) = \bar{\Sigma}(<\theta) - \Sigma(\theta) \equiv
\Sigma_{\rm crit} \gamma_{\rm t}(\theta)\,,
  \label{dsigmaeq}
\end{equation}
where $\Sigma(\theta)$ represents the mass surface density of the lens
at distance $\theta$ from the putative cluster centre, and
$\bar{\Sigma}(<\theta)$ its mean within $\theta$.

The approach that we follow in this work aims at quantifying the weak
lensing mass bias associated with projection effects and its redshift
evolution, depending on the available number density of galaxies from
which we can measure the lensing signal induced by the interposed
cluster projected mass density distribution.  In this way, we quantify
the most optimistic weak lensing mass bias that we expect from \Euclid
data.  In this analysis, we do not assume any uncertainty on the
cluster centre, the lens and source redshifts, all assumed to be
known with infinite accuracy.  Forthcoming analyses, within the Euclid
Collaboration, will be dedicated to study these systematics, and the
corresponding propagation, into the weak lensing mass biases, and concentration that 
should be degenerate and more affected by the miscentring \citep{giocoli21,lesci22}. 

\begin{figure*}
\centering
\includegraphics[width=0.445\hsize]{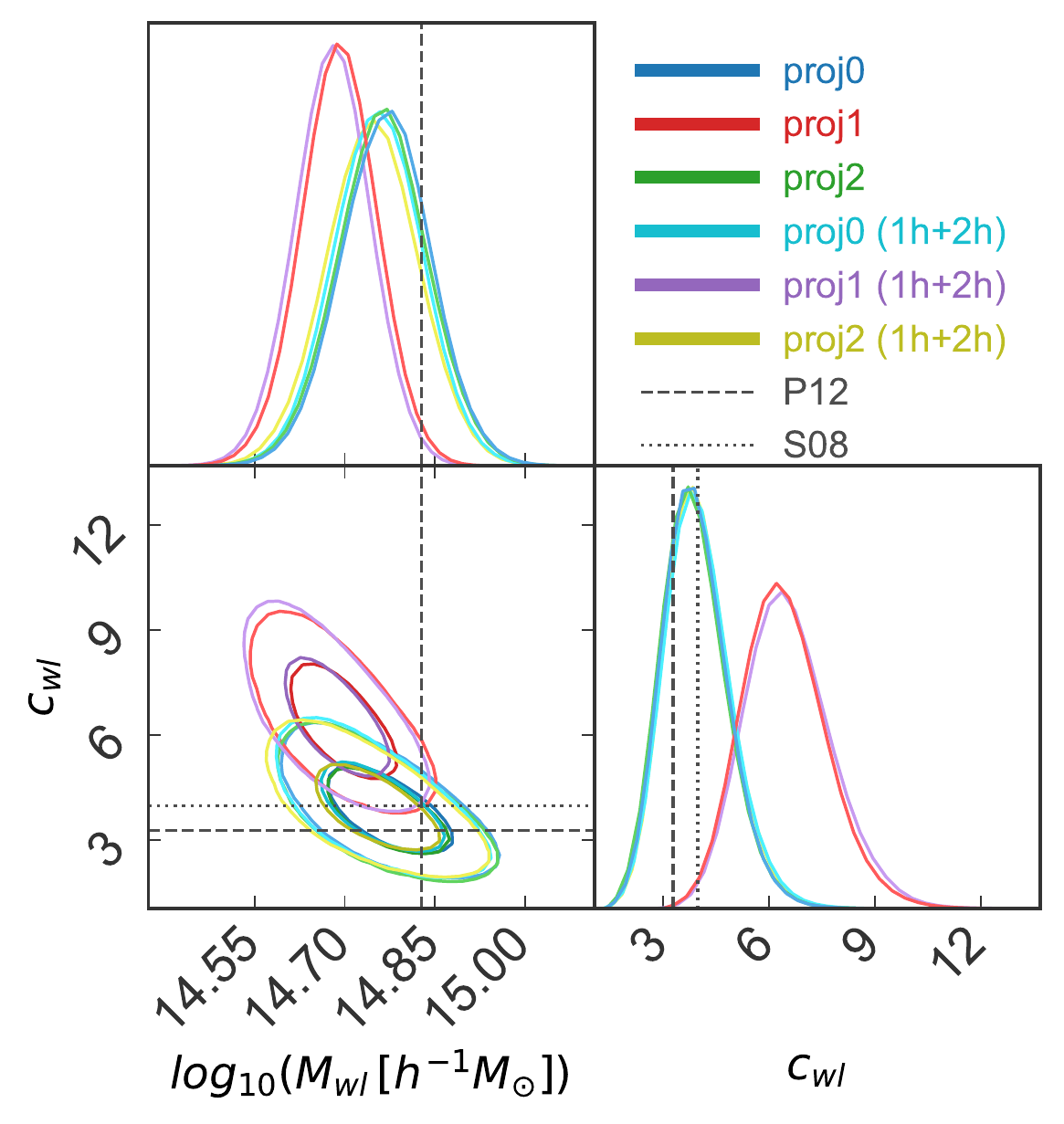}
\includegraphics[width=0.455\hsize]{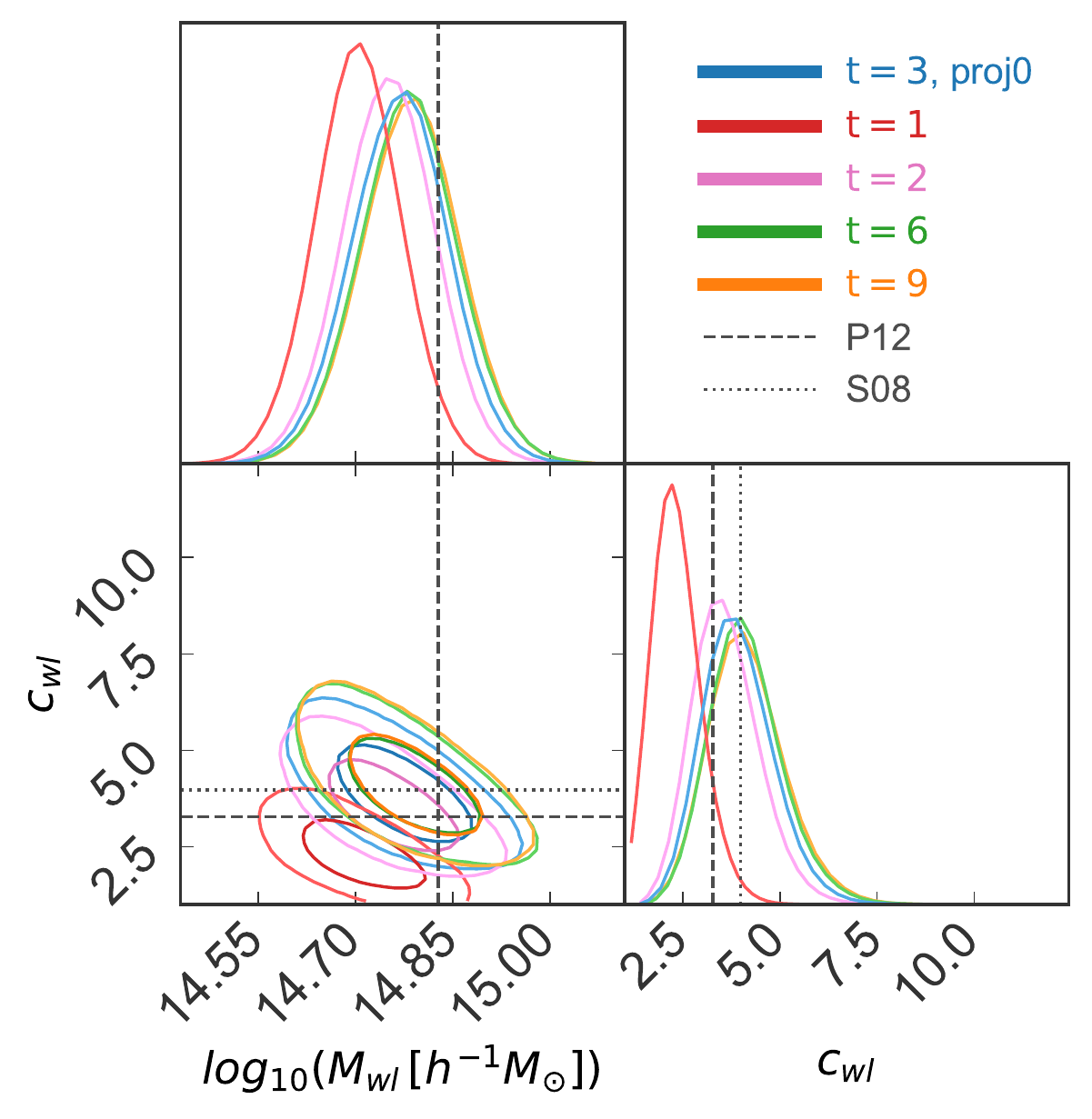}
  \caption{\textit{Left panel}: posterior distributions of the
    recovered logarithm of the mass $M_{\rm wl}$ and concentration $c_{\rm
      wl}$ obtained by fitting the excess surface mass density profile
    of the three random projections of the halo-4 at
    $z=0.22$. \textit{Right panel}: posterior distributions obtained
    modelling the excess surface mass density profile on the halo-4
    when oriented along proj0, assuming different truncation radii. The
    blue data displays the reference case when assuming a truncation
    radius $R_{\rm t}=3\,R_{200}$, while the red, pink, green and
    orange one exhibits the results when using $R_t = t \, R_{200}$
    with $t=1$, 2, 6, and 9,
    respectively. Dark and light-shaded areas enclose $1$ and
    $3\,\sigma$ credibility regions, respectively. In both panels the
    vertical dashed line marks the true mass of the system, while the
    horizontal dotted and dashed ones indicate the concentrations
    computed using the \citet{springel08b} and \citet{prada12}
    formalisms, respectively.\label{fig_chains}}
\end{figure*}

To simulate the weak lensing signal of each cluster, we build the
average excess surface mass density by binning the measured $\Delta
\Sigma (r_i)$, where $r_i \equiv D_{\rm l}\, \theta_i$, in $22$
logarithmically equispaced intervals from 0.02 to
$1.7\,h^{-1}\,\mathrm{Mpc}$ from the cluster centre.  As an example,
in the top panel of Fig.~\ref{figDsigma}, blue, red, and green data
points display the simulated $\Delta \Sigma$ profile for three random
projections of the cluster halo-4 at $z=0.22$, as shown in
Fig. \ref{figkappa}.  In order to limit the analysis to well-constrained shear estimates
 (depending on the angular binning and on the source density), we consider only radial bins with at least 10
simulated data measures. This guarantees a reliable estimate of the average signal and conservatively neglects bins close
to the centre, where baryonic effects tend to steepen the profile. However, we have tested that considering 
radial bins with at least one galaxy does not alter the mass bias results of our work. 
 The colours correspond to those used for the
panel frame of Fig.~\ref{figkappa}. The corresponding error bars are
computed as:
\begin{equation}
\sigma_{\Delta \Sigma} = \sqrt{ \sigma_{ \langle \Delta \Sigma \rangle
  }^2 + \Sigma_{\rm crit}^2 \dfrac{\sigma_{\rm e}^2 }{n_{\rm g} \; \pi
    \left( \theta_2^2 - \theta_1^2 \right) } }\,,
\end{equation}
where $\sigma_{\rm e} = 0.3$
\citep{hoekstra04,hoekstra11,kilbinger14,blanchard20} is the
dispersion of the shape of background source galaxies, and $\theta_1$ and
$\theta_2$ are the lower and the upper bounds of the considered radial annulus. In
our error budget, we also include the error of the mean estimated
excess surface mass density in each radial interval
$\sigma_{\langle\Delta \Sigma\rangle} = \sigma_{\rm rms}/\sqrt{n_{\rm
    g,ring}}$ -- with $\sigma_{\rm rms}$ 
representing the $\Delta \Sigma$ root-mean-square:
\begin{equation}
\sigma_{\rm rms} = \sqrt{\sum_{i=1}^{n_{\rm g, ring}} \dfrac{ \left( \Delta \Sigma_i - \langle \Delta \Sigma \rangle_{\rm ring}  \right)^2}{n_{\rm g, ring}}}\,
\end{equation}
$\langle \Delta \Sigma \rangle_{\rm ring}$ the average value, and  $n_{\rm g,ring}$  the number of
galaxies in each ring -- in order to account for: the projected cluster
triaxiality, correlated large-scale structures, and subhalo
contribution in each annulus \citep{gruen15,gruen11,umetsu20a}. It is
also worth mentioning that this term is negligible with respect to the
one that accounts for the intrinsic shape noise.  From the figure, we
notice that while proj0 and proj2 are similar, proj1 shows a
steepening toward the centre. This highlights that depending on which
projection, structural properties recovered from weak lensing mass and
concentration can be different from each other and even in contrast
with respect to the three-dimensional ones.  
In the figure, we highlight that the data binning of the three projections have a different minimum scale because we exclude the rings with less than 10
galaxies.

\section{Analytical models \& results}
\label{anamodelres}
In order to recover the weak lensing mass and concentration, we model the
data of the excess surface mass density profile using projected
analytical relations.  We use the same strategy that is under
development within the Euclid Consortium in the implementation of the
dedicated processing function: the total density profile is
constructed considering the signal coming from the central part of the
cluster called the 1-halo term, and the one caused by correlated
large-scale structures, named the 2-halo term.  Methods based on
non-parametric formalism
\citep{clowe04a,bradac05,bradac06,diego07,merten09,jauzac12,jullo14,niemiec20}
may give more accurate and precise mass reconstructions, but at the
price of being slow and memory expensive, and in this way, hardly
applicable to the very large statistics of clusters expected from
\Euclid photometric data \citep{sartoris16,adam19}.
 
For the cluster main halo, we adopt a smoothly-truncated NFW density
profile \citep[BMO,][]{baltz09}, defined as:
\begin{equation}
  \rho_{\rm BMO}(r_{\rm 3D}|M_{200},c_{200},R_{\rm t})=\rho_{\rm NFW}(r_{\rm
    3D}|M_{200},c_{200}) \left(\frac{R_{\rm t}^2}{r_{\rm
      3D}^2+R_{\rm t}^2}\right)^2 \,,\label{trunc.NFW}
\end{equation}
with $R_{\rm t} = t\,R_{200}$ with $t$ defined as the truncation
factor. For our reference model, following the results by
\cite{oguri11b}, \cite{bellagamba19}, and \cite{giocoli21} we will
adopt a truncation radius $R_{\rm t}=3\,R_{200}$.  The total mass
enclosed within $R_{200}$, i.e., $M_{200}$, can be thought of as the normalisation of the model and as a mass-proxy of the true enclosed
mass of the dark matter halo hosting the cluster
\citep{giocoli12a}. Writing $r_{\rm 3D}^2$ as the sum in quadrature
between the sky projected coordinate $D_{\rm d} \theta$ and the
line-of-sight $\zeta$ coordinate, and integrating along $\zeta$ we can
write
\begin{equation}
\Sigma_{\rm 1h}(\theta | M_{200},c_{200},R_{\rm t}) = \int_0^{\infty} \rho_{\rm
  BMO}(\theta,\zeta | M_{200},c_{200},R_{\rm t})\, \mathrm{d} \zeta\,.
\label{ptrunc.NFW}
\end{equation}
This projected parametrisation describes the smooth blending of the
halo boundary into the connected large-scale structures, regulated by 
the truncation radius. However, at
larger distances from the cluster centre, the projected lensing signal
starts to increase due to the large-scale structures of the correlated
and the uncorrelated matter along the line of sight: the 2-halo term, describing the asymptotic regime far from the halo centre. 
This term can be written as a function of the integrated linear
matter power spectrum weighted by a Bessel function
\citep{oguri11,oguri11b,sereno17},
\begin{equation}
  \Delta\Sigma_{\rm 2h}(\theta,M_{200}) \!= \!\int_{0}^{\infty}
  \frac{\ell\; \diff \ell}{2\pi}J_2(\ell \theta) \frac{\bar{\rho}_{\rm
      m}(z)\;b_{\rm h}(M_{200};z_{\rm l})}{(1+z)^3D_{\rm l}^2(z)}P_{\rm
    lin,m}(k_{\ell};z_{\rm l})\,,
  \label{surf.dens.2haloterm}
\end{equation}
where $z_{\rm l}$ represents the cluster (lens) redshift, $J_2$ is the
Bessel function of second type, $k_{\ell}=\ell/[(1+z)D_{\rm l}(z)]$
indicates the wave vector mode, $\bar{\rho}_{\rm m}$ is the background
density, $P_{\rm lin,m}$ refers to the linear matter power spectrum
and $b_{\rm h}(M;z)$ is the halo bias, for which we adopt the
\citet{tinker10} model.

The total model can be read as follows:
\begin{equation}
\Delta \Sigma(\theta) = \Delta \Sigma_{\rm 1h}(\theta |
M_{200},c_{200},R_{\rm t}) + \Delta \Sigma_{\rm 2h} (\theta, M_{200})\,.
\end{equation}
It is worth noticing that the 2-halo term contribution at small radii
is expected to be negligible, while its contribution becomes
important, with respect to the 1-halo term, only at scales $\gtrsim
3\,h^{-1}\,\text{Mpc}$ \citep{giocoli21,ingoglia22}.  In their Fig.~1,
\citet{giocoli21} show that for distances between 2 and 3
$h^{-1}\,\text{Mpc}$ from the halo centre the profile is sensitive
also to the truncation radius definition.  Given the size of the
field-of-view of our maps, 3.4$\,h^{-1} \,$Mpc on a side, the 2-halo term has a weak contribution, and we decided to ignore it in our modelling
function.

In the top panel of Fig.\ \ref{figDsigma}, the solid blue curve stands
for the best-fit model to the data points of proj0, associated to the
median values of the posteriors of $\logten(M_{200})$ and
$c_{200}$. The shaded region marks the $1\,\sigma$ uncertainties of
the recovered mass and concentration parameters propagating in the
modelling of the $\Delta \Sigma$ profile, associated with the $16^{\rm
  th}$ and $84^{\rm th}$ percentiles of the posterior
distributions. In the bottom panel, we show the relative difference of
the best-fit models where adopting various values for the truncation
radius with respect to the reference case where we set $R_{\rm t} =
3\,R_{200}$.

The left panel of Fig.\ \ref{fig_chains} displays the posterior
distributions of the recovered mass, $\logten(M_{200})$, and
concentration, $c_{200}$, obtained by modelling the three random
projections of the halo-4 at $z_{\rm l}=0.22$, (top panel of
Fig.~\ref{figDsigma}). In blue, red, and green, we show the
distribution for the projections proj0, proj1, and proj2,
respectively, using only the 1-halo term in our modelling function and
fixing $R_{\rm t}=3\,R_{200}$: we define this as the reference model.

We perform a Monte Carlo Markov Chain run assuming a Gaussian
log-likelihood between the model and the data as defined within our
reference CBL
libraries\footnote{\url{https://gitlab.com/federicomarulli/CosmoBolognaLib}}
\citep{marulli16} that can be read as:
\begin{equation}
\mathcal{L} \propto \exp \left( - \dfrac{1}{2} \chi^2 \right)\,,
\end{equation}
where 
\begin{equation}
\chi^2 = \sum_{i} \left( \dfrac{\Delta \Sigma_{i}(\theta_i) - \Delta
  \Sigma_{\rm model}(\theta_i) }{\sigma_{\Delta \Sigma_i}} \right)^2\,,
  \end{equation}
summing on the number of radial bins.  
Note that we do not account for reduced shear corrections both in the simulations and in the model;
however, it is worth mentioning that observationally this model absorbs the reduced shear 
impact in the resulting mass bias.
We set uniform priors for
$\logten(M_{200}/ [h^{-1}\,M_{\odot}])\in[12.5,\,16]$ and
$c_{200}\in[1,\;15]$, and let each MCMC chain run for $16\,000$ steps.
In cyan, purple, and gold, we show the posterior distributions obtained
by also adding the contribution of the 2-halo term in the modelling
function, as expressed by Eq.~\ref{surf.dens.2haloterm}. Comparing the two, 
we notice that the results are in full agreement with the analysis that does not include the 2-halo term,
also for high-redshift clusters.  This guarantees that for each
cluster and redshift the data-set that we built by projecting the mass density distribution 
from the simulations, takes  a negligible contribution 
from correlated structures along the line-of-sight, due to the limited
size of the field-of-view ($3.4\,h^{-1}\,$Mpc), as discussed earlier.

In the right panel of Fig. \ref{fig_chains}, we examine the dependence
of different assumptions for the truncation radius on the recovered
weak lensing halo mass and concentration, as parametrised in Eq.~\ref{trunc.NFW} and inserted into Eq.~\ref{ptrunc.NFW}. 
We show the posterior
distributions when assuming $t$ to be equal to 1 (red), 2 (pink), 3
(blue), 6 (green), and 9 (orange). From the figure, we notice that the
different choices result in a diverse recovered concentration: smaller
truncation radii prefer lower concentration parameters. When $t=1$
(relatively unphysical case) and 2, the mass also tends to be
underestimated. To guide the reader, we underline that in both panels the dotted and dashed lines refer to the true 3D halo mass and
concentrations, adopting the \citet{springel08b} and \citet{prada12}
relations, respectively.

\begin{figure}
  \includegraphics[width=\hsize]{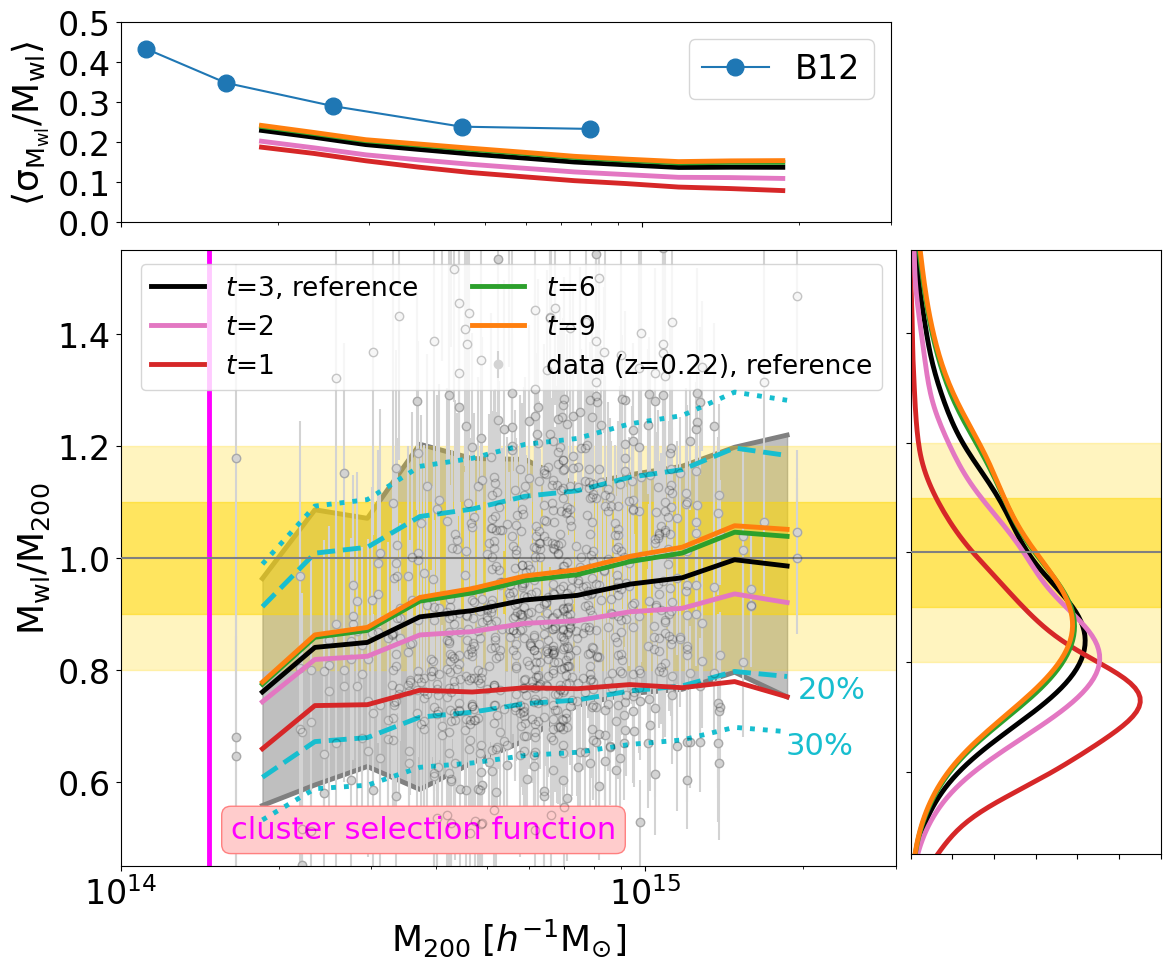}
  \caption{Weak lensing mass bias as a function of the cluster true
    mass, for all systems at $z_{\rm l}=0.22$, for different
    assumptions of the truncation factor.  The data points show the
    ratio between the recovered weak lensing mass obtained by fitting
    the excess surface mass density profile using our reference model
    -- 1-halo term assuming a truncation radius $R_{\rm t}=3\,R_{200}$
    -- and the true $M_{200}$ mass. The error bars correspond to the
    $16^{\rm th}$ and $84^{\rm th}$ percentiles of the posterior
    distribution. The black solid line represents the moving 
    average of the data points. Red, pink, green, and orange lines
    display the moving average of the data considering the
    truncation radius parameter equal to $t=$1, 2, 6, and 9 in our
    modelling function, respectively. The cyan dashed, and dotted lines point to 20 and 30\% of the 
    scatter, respectively, while the magenta vertical line indicates the minimum cluster mass -- almost constant with redshift -- expected to be detected 
    from the photometric catalogue as predicted by \citet{sartoris16}. 
    The light and dark yellow bands
    mark $10\%$ and $20\%$ scatters, respectively.  The top subpanel
    displays the relative weak lensing mass uncertainty as a function
    of the true cluster mass.  The blue circles show the prediction by
    \citet[][hereafter B12]{bahe12}.  The histograms in the right
    panel show the weak lensing mass bias distributions over all
    cluster masses for the various considered cases. \label{figrt119}
  }
\end{figure}

In Fig. \ref{figrt119}, we show the weak lensing mass bias as a
function of the cluster true mass $M_{200}$, for systems at $z_{\rm
  l}=0.22$. The grey data points display the ratio between the weak
lensing mass and the true mass, modelling the excess surface mass
density profile with our reference model: 1-halo term with a
truncation radius parameter $t=3$.  The error bars indicate the
$16^{\rm th}$ and $84^{\rm th}$ percentiles of the posterior
distributions for each considered cluster projection. The black solid
line shows the moving average of the data. Red, pink, green,
and orange lines display the moving average when considering
the truncation radius equal to 1, 2, 6, and 9 times $R_{200}$ in our
modelling function, respectively (the corresponding data points are
not displayed in the figure). From the figure, we notice that the
average weak lensing cluster mass bias depends on the value of the
truncation radius adopted in the modelling. In particular, assuming
$t=1$ we underestimate the mass by approximately $\sim 30\%$, larger
$t$ values increase the weak lensing mass and therefore decrease the
mass bias. For our reference case with $t=3$, the average mass bias is
as low as $\sim 7\%$ with a standard deviation, shown by the grey shaded region \citep[see also ][]{becker11}, 
thus highlighting that triaxiality is the largest source of intrinsic scattering. 
The right panel exhibits the distributions for the corresponding cases independently
of the halo mass $M_{200}$, while the top one shows the average
relative uncertainty for the weak lensing mass estimates as a function
of 3D cluster mass. The blue-filled circles report the results by
\citet[][hereafter B12]{bahe12} that model the weak lensing mass bias
using tangential shear profiles up to 2.2$\,h^{-1}\,$Mpc from the
cluster centre (while we extend our analysis out to 1.7
$\,h^{-1}\,$Mpc).  Their value is larger than ours and the authors are
using only the 1-halo term, modelled with a pure NFW profile (we adopt
a truncated NFW using the BMO relation), without quantifying the
2-halo contribution -- that, in their case, may not be entirely
negligible because of a larger field of view. In addition, they use a
dispersion parameter of the intrinsic shape of background source
galaxies $\sigma_{\rm e}=0.2$ while we adopt a larger value
$\sigma_{\rm e} = 0.3$.  All these factors may increase the relative
uncertainties in the recovered weak lensing masses.

\subsection{Mass bias and concentration}

\begin{figure}
\centering
\includegraphics[width=\hsize]{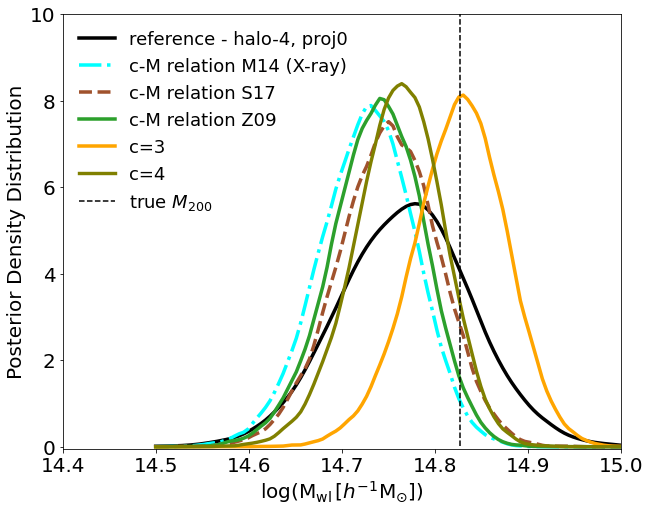}
  \caption{Posterior distributions when recovering the logarithm of
    the cluster mass for proj0 of the halo-4 at $z=0.22$. The black curve shows the reference case where we model both the mass
    and the concentration, as in the top panel of
    Fig.~\ref{fig_chains}. The other histograms show the posteriors
    when assuming a parametrisation of the concentration-mass
    relation; in particular, orange and olive curves refer to the cases
    when assuming a constant value of $c_{200}=3$ and $c_{200}=4$,
    respectively. The vertical black dashed line marks the true value
    of $M_{200}$ as computed by \texttt{AHF}.\label{figDistC}}
\end{figure}

In the spirit of understanding how well we can recover the weak
lensing mass bias when assuming a concentration model, in
Fig.~\ref{figDistC} we show the posterior distribution of the logarithm of the recovered cluster mass in the case of the halo-4 when
considering different assumptions for the concentration. Generally, we
see they have an impact on the recovered mass. In particular, if we
assume a fixed value equal to $c_{200}=3$, we have no bias on the
recovered mass for this specific cluster and projection.
More relevant than mass biases for individual realisations are average biases computed for 
representative cluster samples, 
which show a similar dependence on the adopted concentration mass relation \citep[e.g][]{sommer22}.

\begin{figure}
  \includegraphics[width=\hsize]{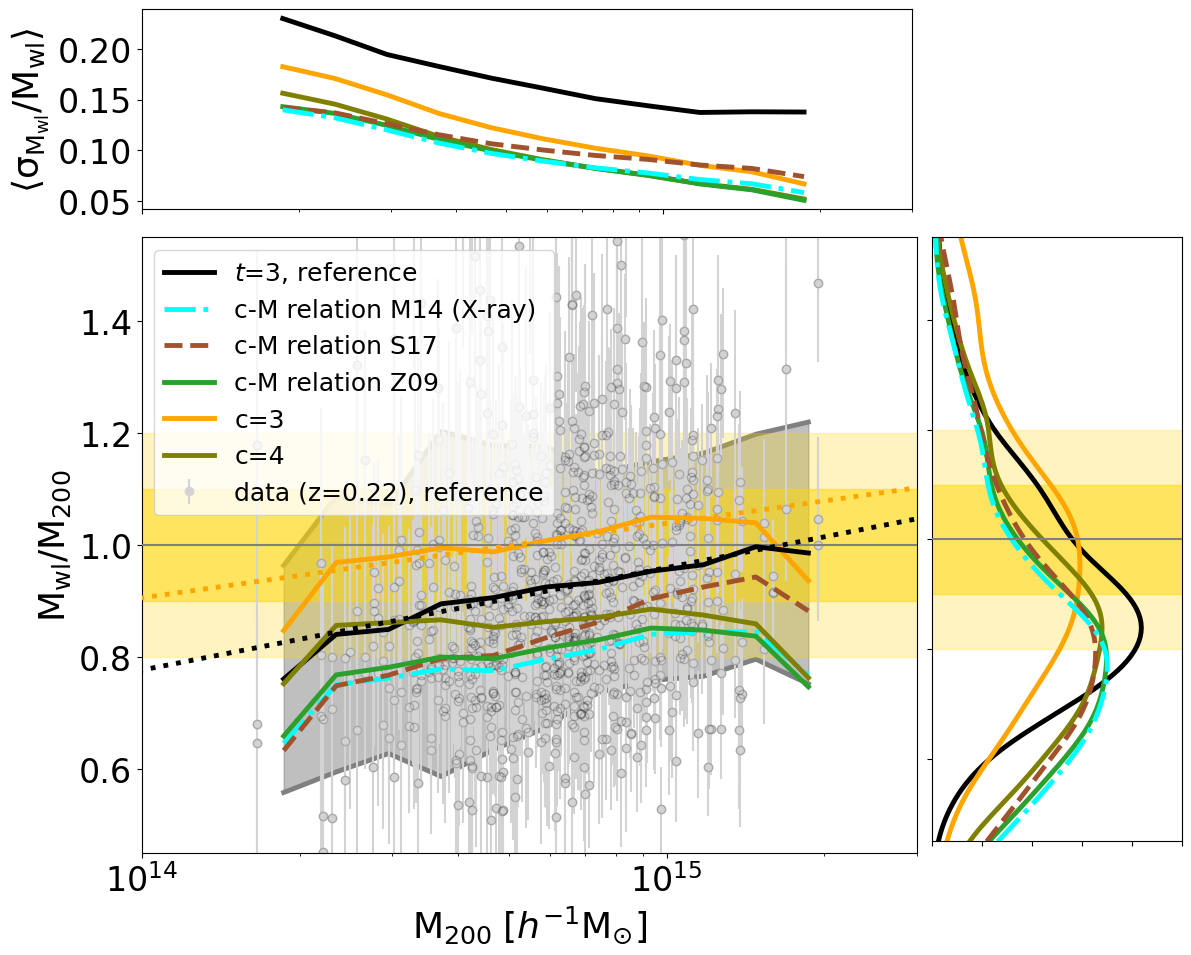}
  \caption{Weak lensing mass biases for clusters at $z=0.22$; the
    points are the same as in Fig.~\ref{figrt119}. The different line
    styles and coloured curves display the moving averages when
    assuming different approaches when modelling the
    concentration. The black solid line refers to the reference case
    where we vary both the mass and the concentration in our MCMC
    analysis. The solid green, dotted cyan, and dark-red curves refer
    to the case in which we assume a concentration-mass relation
    model, as in the figure label. On the other side, the orange and
    olive ones assume a constant value of $c_{200}=3$ and $c_{200}=4$,
    respectively. The coloured dotted lines display the linear regressions
    to the moving average and the corresponding uncertainty (not shown for the orange data to not overcrowd the figure).
    \label{figMbiasC} Top and left panels show the mean
    relative mass uncertainties as a function of the true cluster mass and the distributions of the weak lensing mass biases,
    respectively.}
\end{figure}

The average weak lensing mass bias, as a function of the true cluster
mass, is displayed in Fig.~\ref{figMbiasC}.  The black line shows our
reference case. The solid green, dashed cyan, and dark red lines correspond
to the moving average when assuming different concentration-mass
relation models: \citet{zhao09}, \citet{meneghetti14} and
\citet{sereno17}, respectively. The orange and olive curves display
the results assuming a fixed value $c_{200}=3$ and $c_{200}=4$,
respectively. Referring to the P12 (S08) model the average
concentration of clusters at $z=0.22$ is $\langle c_{200} \rangle=4$
(5) (see the left panel of Fig.~\ref{figconcentration}).
The corresponding coloured dotted lines display the linear regressions:
\begin{equation}
\left\langle\dfrac{M_{\rm wl}}{M_{200}}\right\rangle = \alpha \,\left\langle\log\left(\dfrac{M_{200}}{M_{14.5}}\right)\right\rangle + \beta\,, \label{eqlin}
\end{equation}
where $M_{14.5}=10^{14.5}\,h^{-1}M_{\odot}$, for the reference and $c_{200}=3$ cases (see Tab.~\ref{tab_lin}). The average values are computed assuming a step-size=0.1 and bin-size=0.3
in $\log \left(M_{200}\right)$.

  From the top
subpanel, it is interesting to notice that the average relative
scatter as a function of $M_{200}$ is reduced by approximately a
factor of two when assuming a deterministic model for the
concentration, with respect to the case in which we vary it inside the
Monte Carlo analyses. The right panel displays the distribution of the
relative mass ratio.

\begin{figure*}
  \includegraphics[width=0.445\hsize]{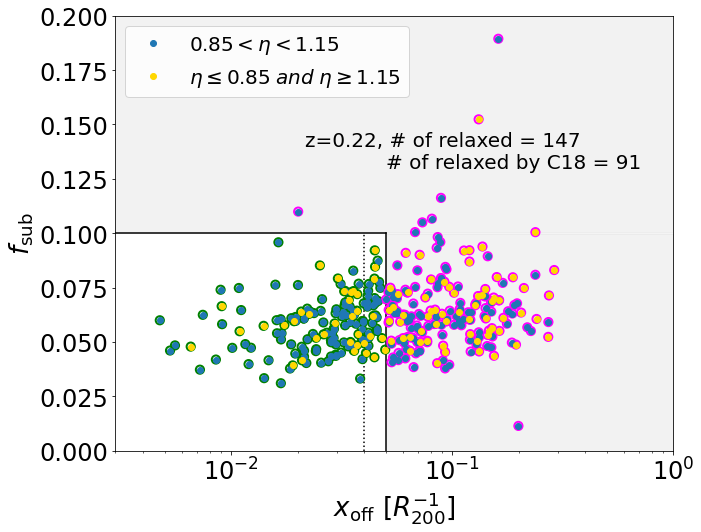}
  \includegraphics[width=0.555\hsize]{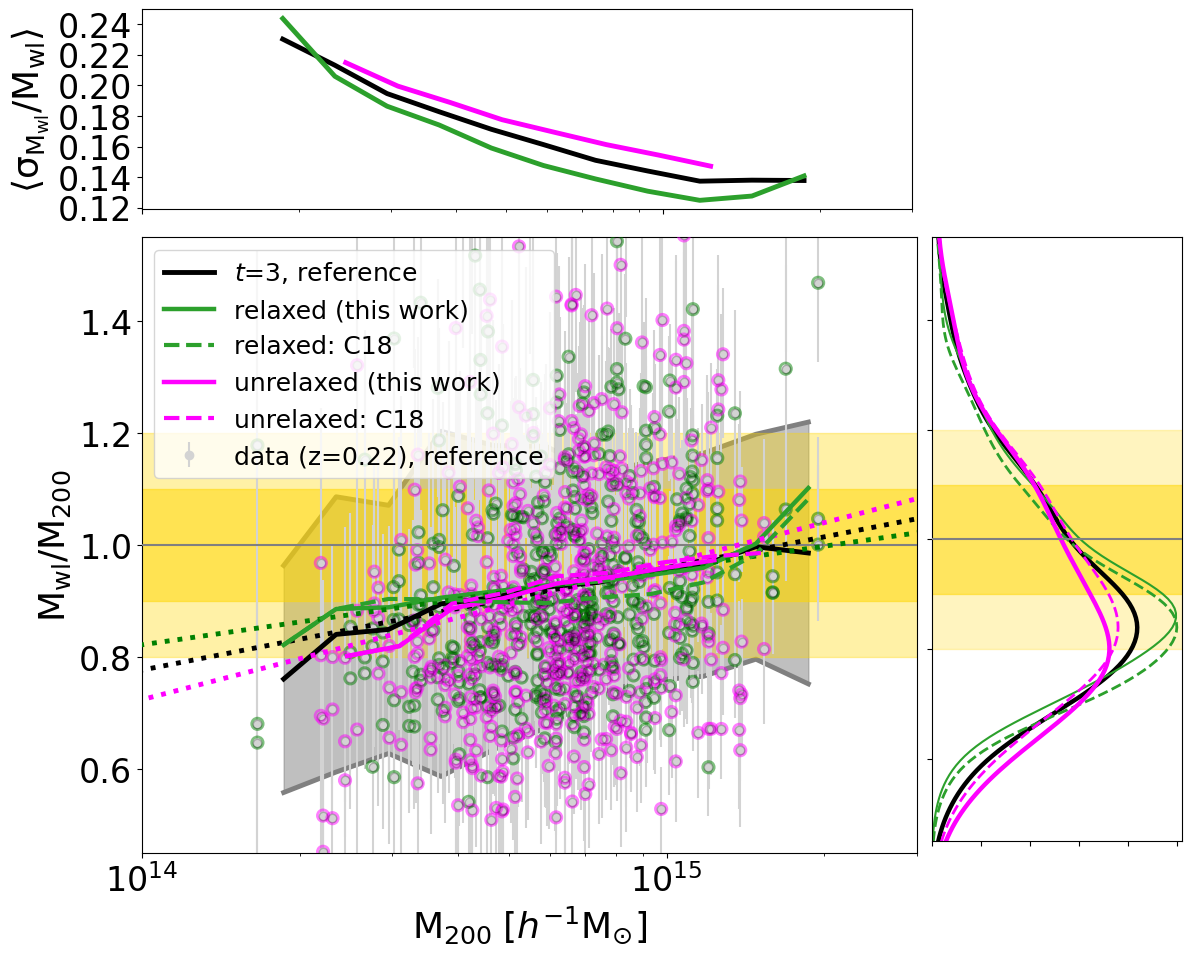}
  \caption{\textit{Left panel}: relaxation criteria for the clusters
    at $z=0.22$. In our analysis, we assume the systems to be relaxed
    if their mass fraction in substructures $f_{\rm sub}<0.1$ and
    their centre of mass offset $x_{\rm off}<0.05\;R_{200}$ indicated by the white
    rectangular region on the bottom left. Our relaxed (unrelaxed)
    clusters are circled in green (purple).  For comparison, we mark
    also the relaxation criterium by \citet{cui18}: $f_{\rm sub}<0.1$,
    $x_{\rm off}<0.04\;R_{200}$ and energy parameter $0.85<\eta<1.15$
    (blue-filled circle, otherwise gold-coloured). \textit{Right
      panel}: weak lensing mass bias as a function of the true mass of
    the cluster. Green and magenta circled data points correspond to
    systems that are or are not relaxed, respectively; the solid green
    and magenta show the moving average of the two
    samples. For comparison, the dashed lines, accordingly coloured,
    show the results using the relaxation criteria by
    \citet{cui18}. The sub-panel on the right displays the mass bias
    distributions for the various cluster samples, while the top one
    is the average relative scatter in the weak lensing mass estimates
    as a function of $M_{200}$.  \label{fig_massbias_relax}}
\end{figure*}

\subsection{Mass bias and relaxation criteria}

The morphological properties of galaxy clusters depend on their mass
assembly history. Typically, clusters that experienced recent merging
events may contain a larger fraction of their mass in substructures
and eventually a mass centroid which is not coincident with the
position of the density peak due to the presence of multi-mass
components.  These properties manifest themselves in a variety of
observables at different wavelengths, going from X-ray to the radio
band.  The projected mass recovered using weak lensing data is also
influenced by the level of dynamical relaxation of the lens. In this
section, we aim at quantifying these dependencies.

In the left panel of Fig.~\ref{fig_massbias_relax} we display the
scatter plot of clusters at $z=0.22$ for parameters $f_{\rm sub}$ and
$x_{\rm off}$, that we use to define the relaxation status of a system
-- see their definition in Sect. \ref{300sims}. We define as relaxed
systems those with $f_{\rm sub}<0.1$ and $x_{\rm off}<0.05\,R_{200}$,
i.e., the clusters lying in the lower left white rectangle of the
figure: 147 over 324 objects satisfy this condition. Our relaxed
clusters are green-circled, while unrelaxed systems are displayed
using magenta circles. In the figure, we also mark the relaxation
criterium adopted by \citet{cui18}: $x_{\rm off}<0.04\,R_{200}$ (with
a dotted vertical line) and with virial ratio $0.85<\eta<1.15$. The
latter condition is portrayed in the figure, colouring the data points
in blue or otherwise in yellow. The relation criterion adopted by
\citet{cui18} is satisfied by 91 systems.  In the right panel, we
display the weak lensing mass bias, as a function of the cluster mass,
for the relaxed (green) and the unrelaxed (magenta) clusters. The
green and magenta solid curves represent the moving  average
of the two samples, respectively. The corresponding dashed curves
refer to the cases in which the relaxation criterion by \citet{cui18}
is adopted, which also uses information on the virial energy
ratio. From the top subpanel, we notice that the average relative
uncertainty in the recovered mass is lower for relaxed systems -- even
if the differences between the two cases are, to some extent, small,
as well as the relative mass bias has a lower scatter than for
unrelaxed clusters (right subpanel). This highlights the fact that
unrelaxed systems could not be well modelled by the single halo model
\citep{lee23}. As in previous figures, and to highlight the differences between the relaxed and
unrelaxed samples, the coloured dotted lines display the linear regressions
to the moving average and the corresponding uncertainty. The best-fit parameters and the corresponding uncertainties, 
as expressed in Eq.~\eqref{eqlin} for the 
relaxed and unrelaxed sample at $z=0.22$ are shown in Tab.~\ref{tabrelaxed}, which can be compared with the full sample as in Tab.~\ref{tab_lin}.

\begin{table}[h]  
\begin{center}
\caption{Linear regression parameters, as in Eq.~\eqref{eqlin}, for the relaxed and unrelaxed cluster sample at $z=0.22$. \label{tabrelaxed}}
\begin{tabular}{lcc}
\hline
    & Slope $\alpha$  & Intercept $\beta$ \\
\hline \hline
relaxed   &   $ 0.1354 \pm 0.029 $ &    $0.890\pm 0.596 $    \\
unrelaxed   &  $ 0.243 \pm 0.033 $ &    $0.846\pm 0.681 $    \\ 
\hline
\end{tabular}
\end{center}
\end{table}

\subsection{Mass bias along preferential directions}

\begin{figure*}
  \includegraphics[width=0.441\hsize]{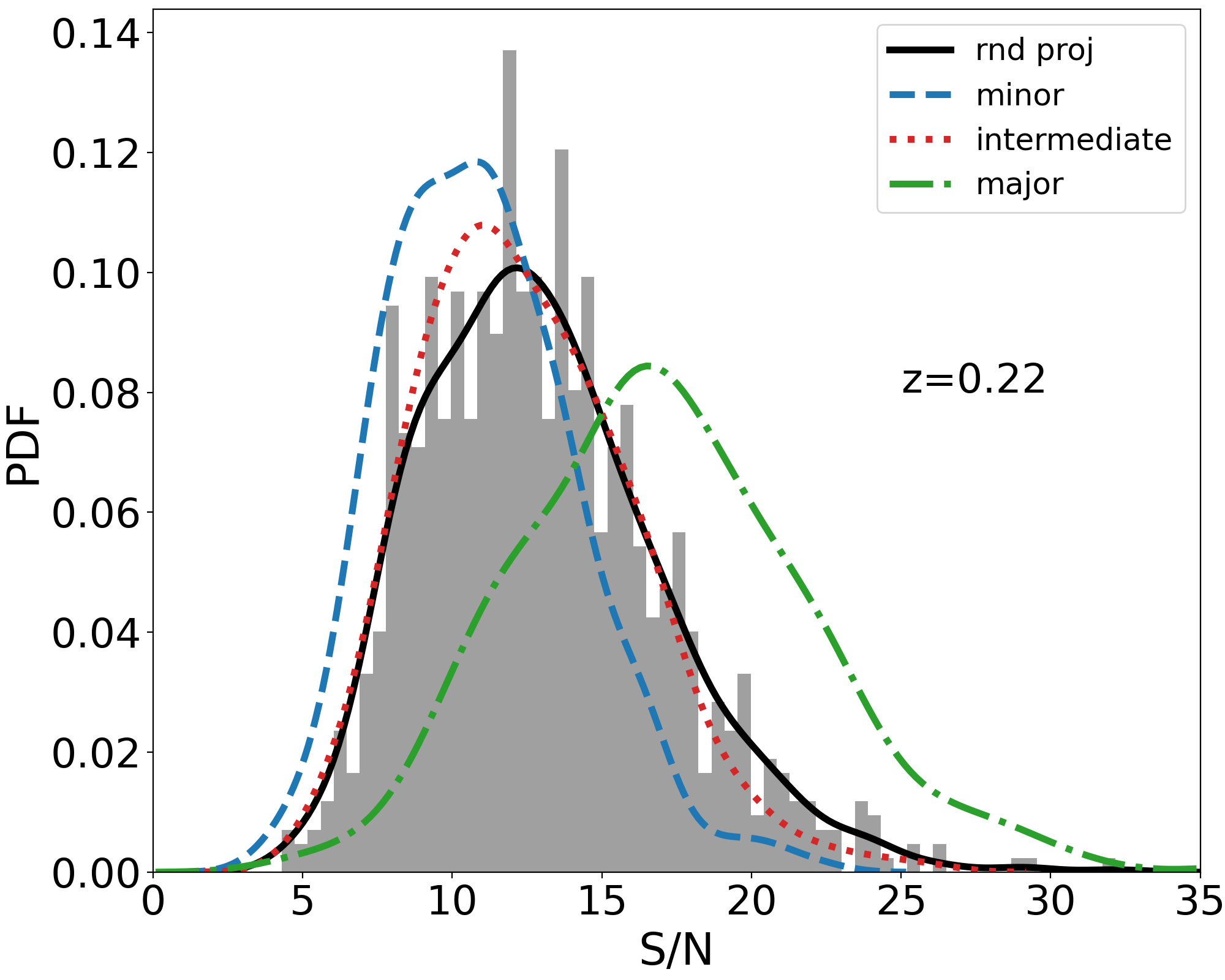}
  \includegraphics[width=0.559\hsize]{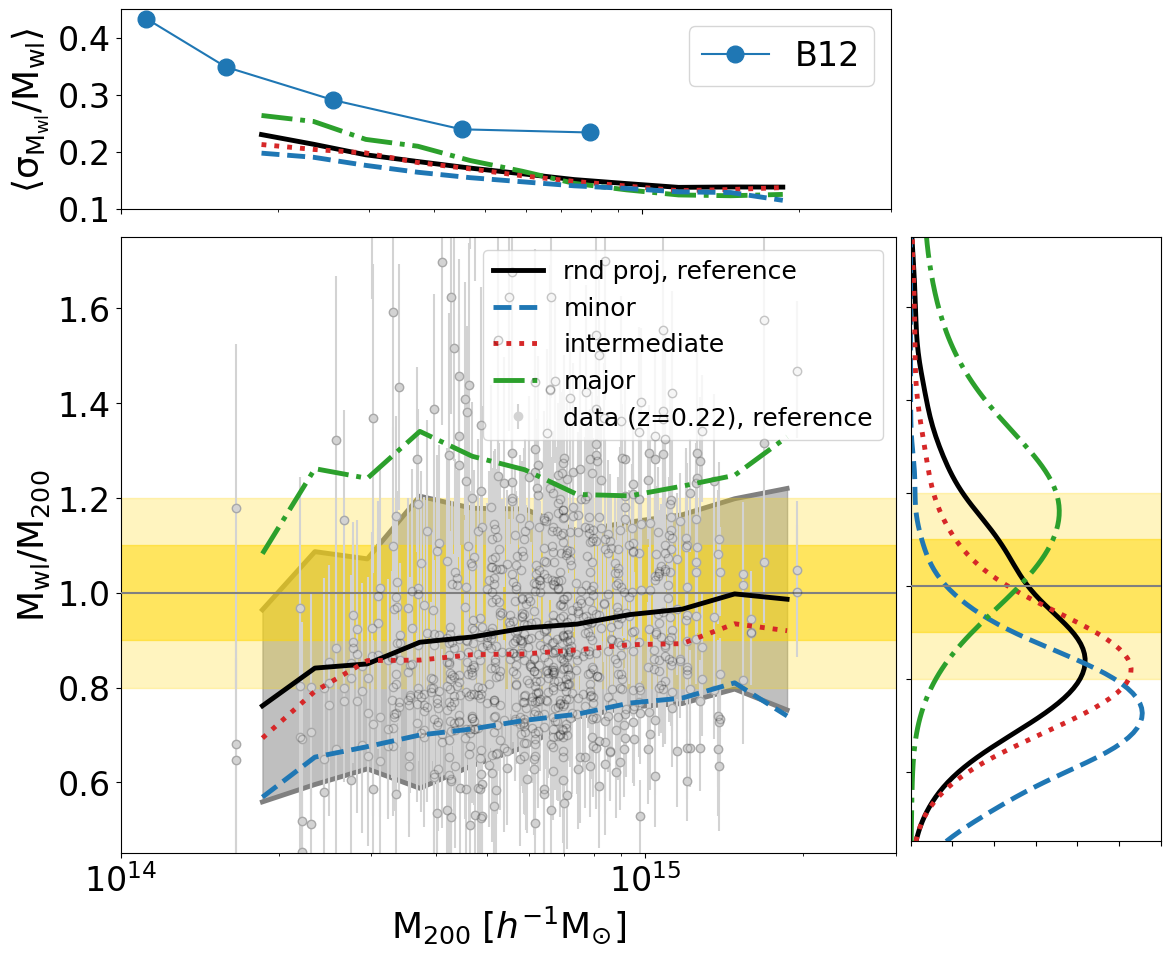}
  \caption{\textit{Left panel}: weak lensing signal-to-noise
    distributions for the sample of clusters at $z=0.22$. The grey-shaded histogram shows the distribution for the clusters oriented
    along random projections, namely the $x$, $y$, and $z$ cartesian
    axes of the re-stimulated box region. The solid black curve
    represents the Kernel Density Estimate (KDE) of the grey discrete
    histogram. The blue (dashed), red (dotted), and green (dot-dashed) curves
    refer to the KDE of the distributions when systems are oriented
    along the minor, intermediate, and major axis of the cluster
    ellipsoid, respectively. \textit{Right panel}: weak lensing
    derived mass bias as a function of the true mass of the cluster.
    Blue (dashed), red (dotted), and green (dot-dashed) solid lines
    display the moving average of the corresponding mass bias
    data points of the preferential projections, minor, intermediate,
    and major axis, respectively.\label{figorientationbias} The top
    and right sub-panels display the relative mass uncertainties as a
    function of the true mass, and weak lensing mass bias
    distributions, respectively.}
\end{figure*}

\begin{figure}
  \includegraphics[width=\hsize]{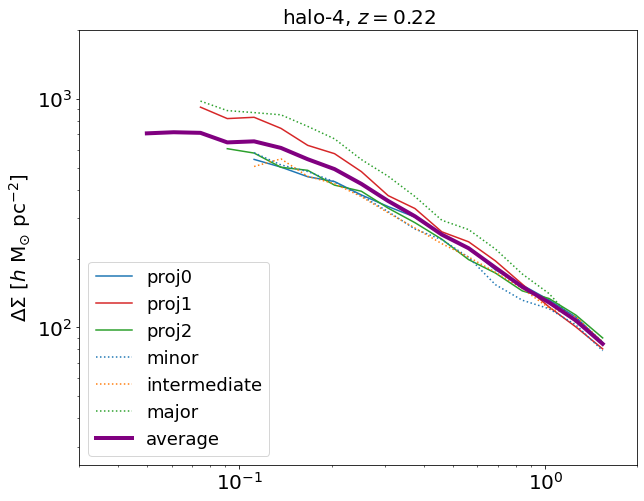}  
  \includegraphics[width=\hsize]{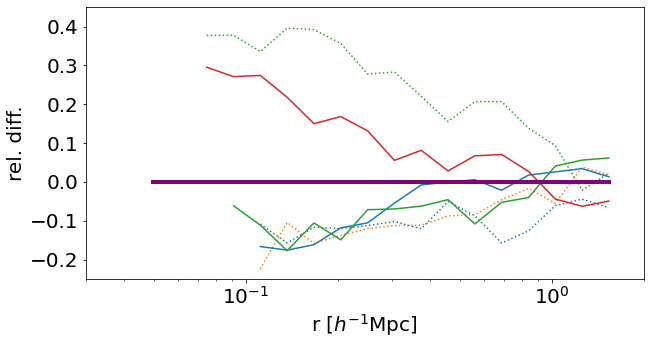}  
  \caption{\label{figstacked} \textit{Top panel}: average excess
    surface mass density profile averaging six different projections of
    the same cluster (halo-4 at $z=0.22$): three randoms and three preferentials.  For comparison, solid blue, red, and green curves show the
    individual profiles around the three random projections, while
    dashed blue, orange and green when the cluster is oriented along
    the minor, intermediate and major axis of the moment of inertia
    tensor ellipsoid.  \textit{Bottom panel}: relative difference of
    the individual profiles with respect to the averaged one.}
\end{figure}

The reliability with which the cluster mass can be recovered is also
related to the orientation of the moment of inertia tensor ellipsoid
with respect to the line-of-sight
\citep{sereno12,sereno13,despali14,despali17,sereno18b}.
\citet{herbonnet22} have shown that the shape of the bright central galaxy can give
unbiased information about the shape of the total halo mass ellipsoid
and that the recovered mass via weak gravitational lensing is biased
low (high) if the halo is oriented along the minor (major) axis with
respect to the observer.  In this section, we quantify in detail how
the mass bias depends on the orientation of the mass tensor ellipsoid
with respect to the line of sight. Following \citet{sereno17} and
\cite{umetsu20a}, we can define the weak lensing signal-to-noise ratio
as:
\begin{equation}
\dfrac{S}{N} = \dfrac{ \dfrac{\sum_i \Delta \Sigma_i \, \sigma_{\Delta
      \Sigma_i}^{-2}}{\sum_i \sigma_{\Delta \Sigma_i}^{-2}} }{
  \left( \sum_i  \sigma_{\Delta \Sigma_i}^{-2} \right)^{-1/2} } \,,
\end{equation}
where the variable $i$ runs on all the considered radial bins.

In the left panel of Fig. \ref{figorientationbias}, we show the weak
lensing signal-to-noise ratio of clusters at $z=0.22$. The grey
histogram shows the distribution of the $S/N$ for all considered
random projections, while the black curve displays the Kernel Density
Estimate (KDE) of the binned histogram. Blue (dashed), red (dotted), and
green (dot-dashed) curves refer to the distributions of the weak lensing 
signal-to-noise ratios when cluster ellipsoids are oriented with
respect to the minor, intermediate, and major axis of the mass tensor
ellipsoid.  The triaxial parameters used are those computed by the AHF
algorithm; we refer to \citet{knollmann09} and \citet{cui18} for more
details.  The results confirm that orientation matters: weak lensing
signal-to-noise ratio distributions are shifted toward larger (lower)
values when clusters are oriented along the major (minor) axis of the
mass tensor ellipsoid with the line of sight.  Projection effects
also impact the weak lensing mass bias and the scatter, as we can observe in the right
panel of Fig. \ref{figorientationbias}.  The mass bias, if clusters are preferentially selected either along the line of sight or in the plane of the sky, 
 is approximately $+25\%$ or $-25\%$. From the top subpanel, we can also notice that the average
relative weak lensing mass uncertainty varies with the cluster
ellipsoid orientation.

Before concluding this section, it is worth devoting a few words to
cluster detections and orientation biases.  At fixed mass, when
systems are oriented along the major axis of the ellipsoid, they tend
to have, on average, a more compact and concentrated galaxy satellite
distribution than when they are oriented along the minor axis: this
indicates that optical cluster finder algorithms in observational data may
suffer from orientation bias \citep{wu22}.  We plan to examine and
analyse this argument in more detail in a future dedicated work.

\begin{figure}
  \includegraphics[width=\hsize]{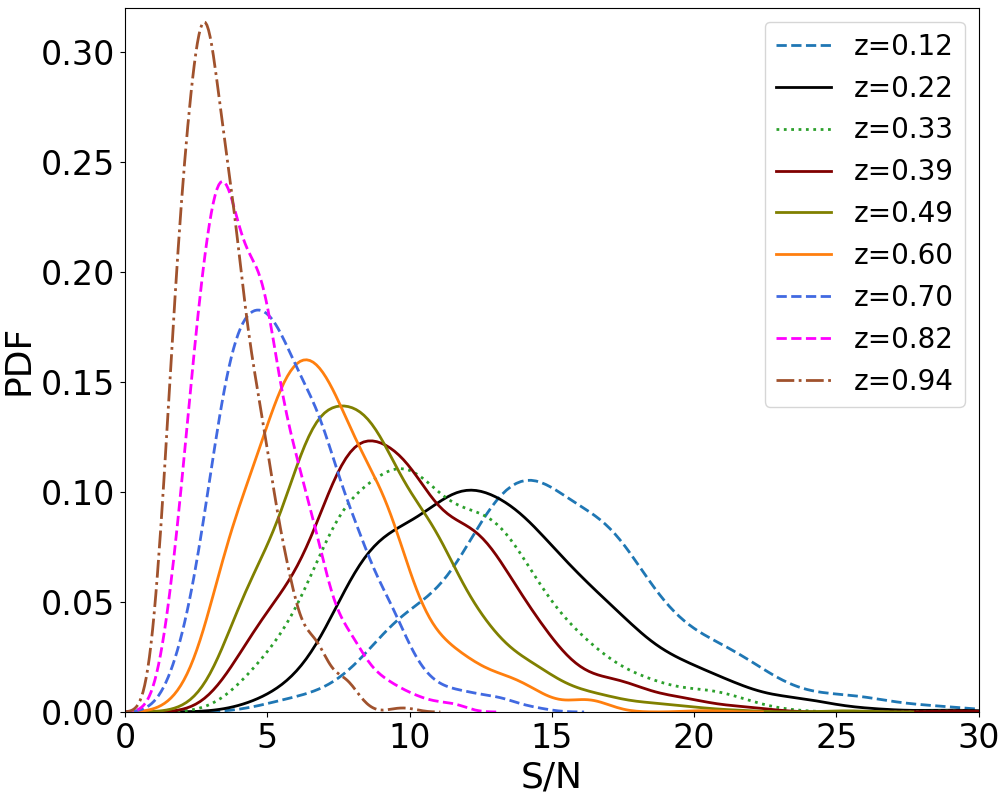}
  \caption{Individual cluster weak lensing signal-to-noise ratio distributions, for random projections and different redshifts, as
    expected in the \Euclid wide-field survey of a cluster population that is representative of the simulated clusters studied here. \label{PDFsnr}}
\end{figure}

For example, in Fig.~\ref{figstacked}, the magenta solid curve shows 
the average excess surface mass density profile of
the 6 projections of cluster halo-4 at redshift $z=0.22$.   
 Solid red, blue, and green lines are the individual profiles for three random
projections. Dotted blue, orange, and green curves show the profiles
when the cluster is oriented along the minor, intermediate, and major
axis of the ellipsoid, with respect to the line of sight, as done
previously.  Only radial bins with more than 10 galaxies are
shown. 


\begin{figure}
  \includegraphics[width=\hsize]{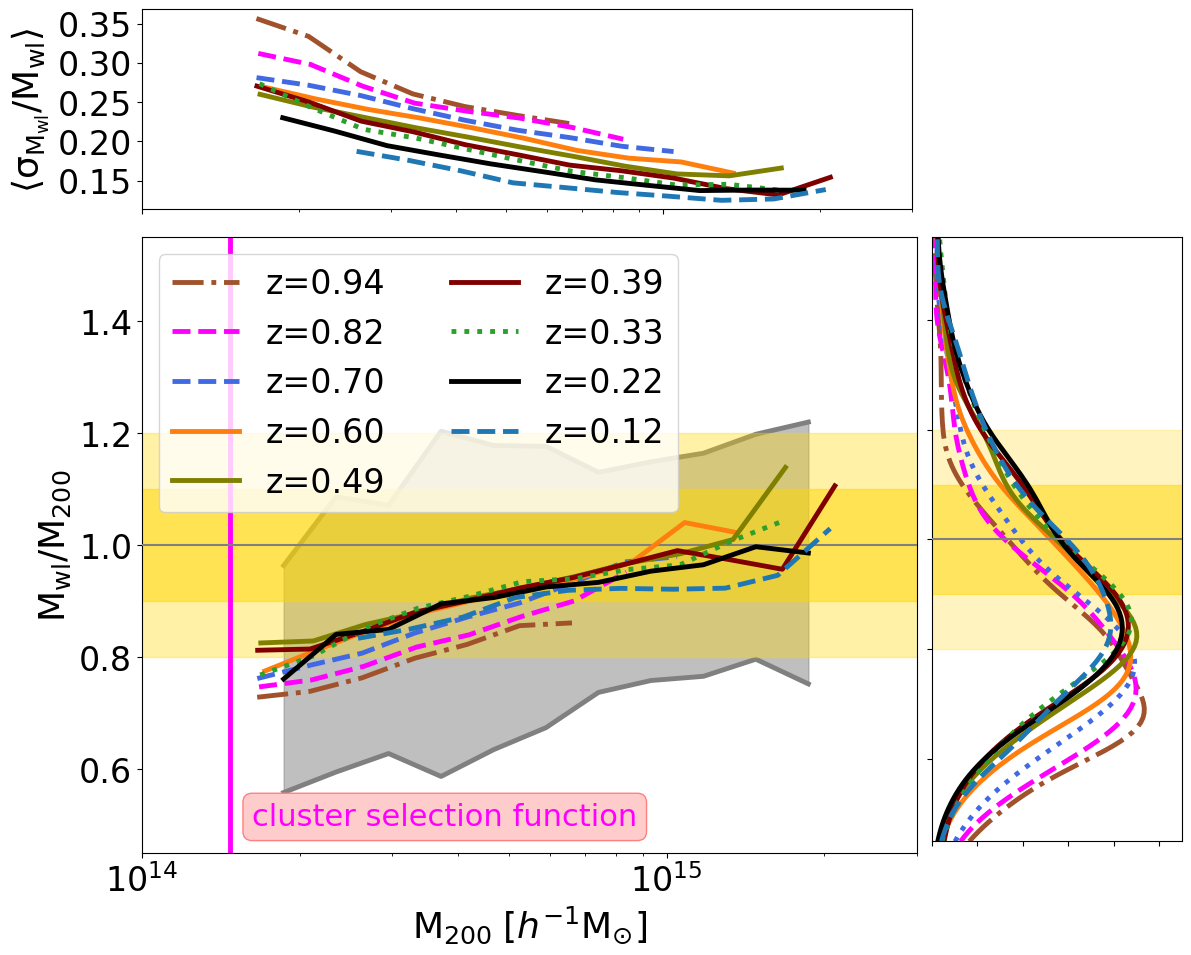}
  \caption{Average weak lensing mass bias as a function of the cluster
    halo mass $M_{200}$ for different lens redshifts. 
     We consider all clusters with $M_{200}>10^{14}M_{\odot}$, above the 
    minimum mass expected to be selected in the photometric catalogue \citep{sartoris16}, relatively constant with redshift.
    Right and top
    subpanels are the same as
    Fig. \ref{figrt119}.\label{figmassbiasz}}
\end{figure}

\subsection{Reshift evolution of the mass bias}

The expected source redshift distribution available for weak lensing measurements tells us how many background sources can be used to probe
the total projected mass distribution of a cluster acting as a
gravitational lens. The reliability of the lensing measurement also
depends on the intrinsic magnitude of the sources and their projected
distance with respect to the cluster centre because of possible 
confusion with cluster member galaxies.  The weak lensing
signal-to-noise ratio of individual clusters is also modulated by the
expected available number density of sources expected from the \Euclid wide survey, as we display in Fig.~\ref{PDFsnr}. Low-redshift systems
have a larger weak lensing signal-to-noise ratio than the high-redshift ones. In the figure, we consider only the measurement of
the three random projections per cluster since we have already
discussed in the previous section that the three preferential ones
need particular attention. We also underline that eventual stacking procedures 
will shift the expected weak lensing signal-to-noise ratio distributions 
toward larger values by a factor that is approximately equal to 
the square root of the number of systems that are combined.

In Fig.~\ref{figmassbiasz} we exhibit the average weak lensing mass
bias for the different redshifts, considering the three random
projections per cluster,  and cluster masses with $M_{200}>10^{14}M_{\odot}$. 
In our modelling analysis, we assume a
uniform range both for the logarithm of the mass and for the
concentration parameter. On average, higher redshift clusters tend to
be more biased low with respect to $M_{200}$, as a result of their
lower signal-to-noise ratio.  Looking to the top subpanel, it is
interesting to notice that the relative error of the recovered weak
lensing mass tends to be larger for systems at higher redshifts, which is 
due to the lower background source densities and average lensing
efficiencies for those systems. In Tab.~\ref{tab_lin} we summarise the values of the parameters 
describing the linear fitting functions and the corresponding  
uncertainties, as in Eq.~\eqref{eqlin},  at different redshifts, for the reference modelling case, and assuming a constant concentration $c_{200}=3$.

\begin{figure}
  \includegraphics[width=\hsize]{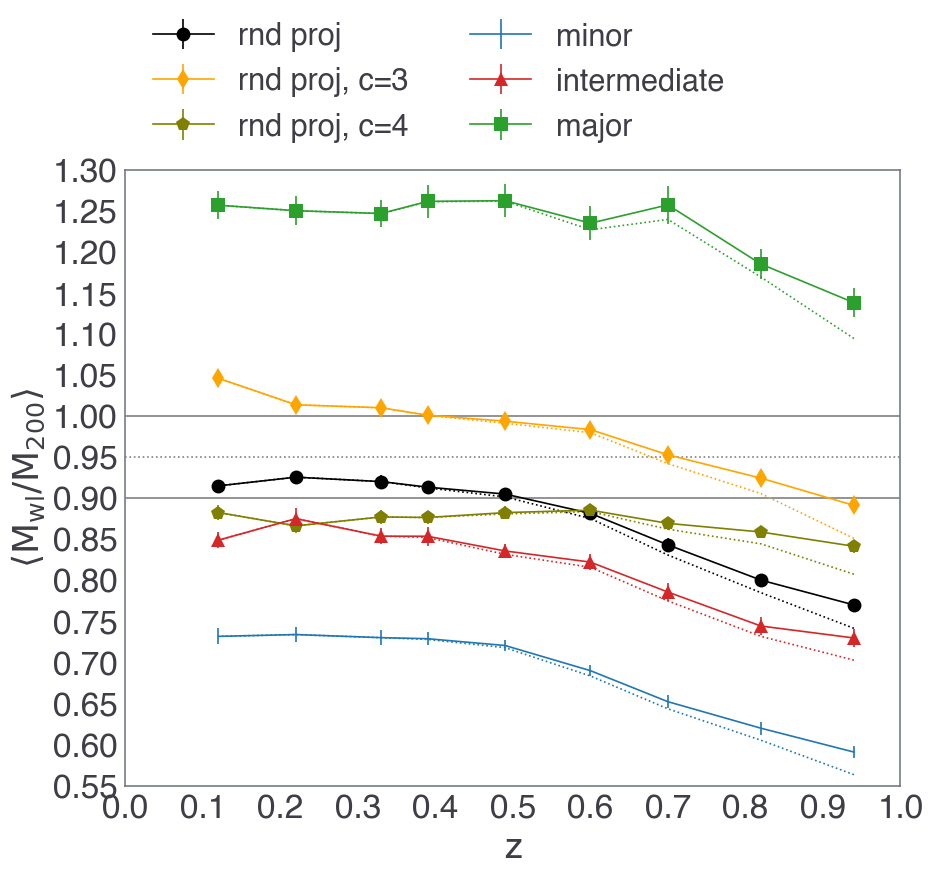}  
  \caption{Average cluster weak lensing mass bias as a function of the
    lens redshift, for clusters with $M_{200}>10^{14}M_{\odot}$. The various data points and colours refer to
    different ways of computing the cluster masses. Black circles
    display the case of random projections and modelling both the halo
    mass and concentration, orange diamonds, and olive-coloured
    pentagons consider the cases in which we assume a fixed
    concentration parameter -- three and four, respectively -- and we
    fit only the halo mass. Light-blue crosses, red triangles and
    green squares show the average mass bias for the three particular projections, along the minor, the intermediate, and the major
    axis, respectively. The dotted lines, 
    which slightly deviated from the data points only at $z>0.6$, correspond to the results for the whole cluster sample, with 
    no minimum mass cut.
    \label{zmassbiasfig}}
\end{figure}

Averaging over all cluster masses with $M_{200}>10^{14}M_{\odot}$, we display in
Fig.~\ref{zmassbiasfig} the weak lensing mass bias as a function of
the lens redshift. Black circles show the results averaged over all
random projections. We can notice that higher redshift systems are
more biased low with respect to the low redshift ones. Blue crosses
and green squares represent the redshift evolution of the mass bias when the clusters are oriented along the minor or major axis of the
ellipsoid with respect to the line of sight, respectively. 
Orange diamonds and olive pentagons refer to the case when in individual random projections, we model only the logarithm of
the cluster mass while keeping the concentration fixed to $c_{200}=3$
and $c_{200}=4$, respectively.  Those values represent the typical
concentration parameters of cluster size-halo as highlighted in
different numerical simulation analyses
\citep{zhao09,giocoli12b,ludlow12,ludlow16}.  Assuming a fixed
concentration $c_{200}=3$ in our modelling analysis of the weak
lensing signal of individual clusters, we can recover, on average, a
weak lensing mass that is biased by no more than 5\% with respect to
the true one, up to redshift $z=0.7$ -- positive for $z<0.4$ and
negative at higher redshift.  The error bars associated with all data
points correspond to the errors on the mean value. The dotted lines, 
which differ more significantly from each other at $z>0.6$, display the results for the whole cluster sample with no minimum mass cut.

\begin{table}[h]  
\begin{center}
\caption{Linear regression parameters as in Eq.~\eqref{eqlin}, at different redshifts for the reference and $c_{200}=3$ cases, 
considering random projections and for clusters more massive than $M_{200}>10^{14}M_{\odot}.$}
\label{tab_lin}
\begin{tabular}{lccc}
\hline
snap.  &  $z$ & Slope $\alpha$  & Intercept $\beta$  \\
\hline \hline
reference   &     0.12   & $ 0.090 \pm 0.028 $ &    $0.883\pm 0.598 $    \\
$c_{200}=3$  &     " & $-0.017\pm 0.035$ & $1.053\pm 0.736$      \\ \hline 

   &     0.22   & $ 0.183 \pm 0.022 $ &    $0.869\pm 0.452 $    \\
  &     "  & $0.133\pm 0.027$  &  $0.973\pm 0.567$   \\ \hline

   &     0.33   & $ 0.208 \pm 0.023 $ &    $0.874\pm 0.473 $    \\
  &     "  &  $0.196\pm0.025$  &  $0.966\pm0.524$  \\ \hline

   &     0.39   & $ 0.215 \pm 0.022 $ &    $0.873\pm 0.464 $    \\
  &     "  & $0.142\pm0.027$ &   $0.974\pm0.549$   \\ \hline
  
     &     0.49   & $ 0.223 \pm 0.022 $ &   $0.874\pm 0.453 $    \\
  &     "   & $0.064\pm 0.024$ & $0.985\pm0.501$    \\ \hline

     &     0.59   & $ 0.273 \pm 0.019 $ &    $0.862\pm 0.401 $    \\
  &     "  & $0.126\pm0.024$ &  $0.976\pm 0.486$    \\ \hline

     &     0.70   & $ 0.292 \pm 0.024 $ &    $0.836\pm 0.488 $    \\
  &     "  & $0.172\pm 0.027$ &  $0.951\pm 0.556$     \\ \hline

     &     0.82  & $ 0.279 \pm 0.023 $  &    $0.811\pm 0.473 $    \\
  &     "  & $0.218\pm 0.028$  &   $0.935 \pm 0.564$  \\ \hline

     &     0.94  & $ 0.255 \pm 0.024 $  &    $0.790\pm 0.498 $    \\
  &     "  & $0.314 \pm 0.032$ & $0.920 \pm 0.653$     \\ \hline
\hline
\end{tabular}
\end{center}
\end{table}

The impact of the scatter and the redshift evolution of the cluster mass bias will be 
of utter importance for cluster cosmology. Different future works have already been 
planned and organised to further assess systematics and nuisance parameters in the 
cosmological likelihood pipeline. 

\section{Summary \& Conclusions}
\label{summary}
In this work, we have performed a systematic study of the weak lensing
mass bias using hydrodynamical simulations of clusters.  We have
simulated the expected excess surface mass density profile from the expected number density of background sources of the ESA \Euclid
wide-field survey, normalised to 30 galaxies per square arcminute. We
have adopted a projected truncated NFW profile to model the data.  Our
main results are summarised as follows:
\begin{itemize}
\item on average individual weak lensing masses $M_{\rm wl}$ are typically lower by
  $5\%$ than the true one, various projections of the same cluster may
  have different recovered weak lensing masses differing by up to  $30\%$: cluster triaxiality represents the largest 
  source of intrinsic scattering;
\item in our reference model, we have adopted a truncation radius of
  three times $R_{200}$, a lower truncation radius returns a more
  biased weak-lensing mass: from $t=1$ to $t=9$ the mass bias for
  clusters at $z=0.22$ ranges from $-30\%$ to a few percent;
\item in modelling the data set we have also investigated the impact
  of using a concentration both dependent, in one case, and
  independent of the total mass in another case, finding that those
  systematically impact the recovered weak lensing mass \citep[see e.g. also][]{sommer22};
\item relaxed clusters, being better described by a 1-halo projected
  density profile are less biased than the unrelaxed ones by
  approximately $3\%$;
\item clusters oriented along the major (minor) axis of the mass
  tensor ellipsoid have their weak lensing mass overestimated
  (underestimated) by approximately $+(-)$ $25\%$;
\item averaging over all masses, when also varying the concentration,
  the negative mass bias tends to increase as a function of redshift;
  however, assuming a fixed concentration, this effect is reduced.
The increase of the negative mass bias with redshift is due to the reduced $S/N$ 
of the weak lensing mass constraints of high-redshift clusters.
As demonstrated by \citet{sommer22} this $S/N$ dependence can be avoided 
if the weak lensing mass scatter is separated into an intrinsic component 
(e.g.~due to variations in density profiles and substructures) and a shape noise component \citep[e.g.][]{bocquet19,chiu22}.	  
  
\end{itemize} 

The use of galaxy clusters as cosmological probes relies on the
accuracy and precision of mass measurements.  The future ESA \Euclid
mission will make use of weak gravitational lensing to determine the
projected total mass.  In this work, we have quantified the accuracy
and bias effects by using dedicated lensing simulations of clusters in
a variety of dynamical states and at different redshifts, based on a
specific set of hydrodynamical simulations: the Three Hundred
clusters.  While the mass bias depends on the considered choice of the
modelling function, the scatter is driven by the complex triaxial
structure of galaxy clusters. However, assuming a fixed value $c_{200}=3$
for the concentration parameter, when modelling the data of individual
clusters, produces less unbiased results, at least for the set of
simulated clusters analysed in this work.

We would like to conclude underling that future works will be devoted to study whether 
these  calibrations are robust against variations in the astrophysical
processes included in the simulations, paving the way for future
cluster analyses and opening new perspectives for the cosmological
exploitation of galaxy clusters.

\begin{acknowledgements}
CG, MM, and LM acknowledge support from the grant PRIN-MIUR 2017
WSCC32 ZOOMING, and the support from the grant ASI n.2018-23-HH.0.  CG
acknowledges funding from the Italian National Institute of
Astrophysics under the grant "Bando PrIN 2019", PI: Viola Allevato and
from the HPC-Europa3 Transnational Access programme HPC17VDILO.  GC
also thanks the support from INAF theory Grant 2022: Illuminating Dark
Matter using Weak Lensing by Cluster Satellites, PI: Carlo Giocoli.
MM acknowledge financial support from PRIN-MIUR grant and 2020SKSTHZ,
INAF ``main-stream'' 1.05.01.86.20: "Deep and wide view of galaxy
clusters (P.I.: M. Nonino)" and INAF ``main-stream'' 1.05.01.86.31
"The deepest view of high-redshift galaxies and globular cluster
precursors in the early Universe" (P.I.: E. Vanzella). SB and MM
acknowledge financial support from the InDark INFN Grant.  GY and AK
would like to thank the Ministerio de Ciencia e Innovaci\'on (Spain)
for financial support under research grant PID2021-122603NB-C21 The
THREE HUNDRED simulations used in this work have been performed in the
MareNostrum Supercomputer at the Barcelona Supercomputing Center,
thanks to CPU time granted by the Red Espa\~nola de
Supercomputaci\'on.  WC is supported by the STFC AGP Grant
ST/V000594/1 and the Atracci\'{o}n de Talento Contract
no. 2020-T1/TIC-19882 granted by the Comunidad de Madrid in Spain. He
also thanks the Ministerio de Ciencia e Innovación (Spain) for
financial support under Project grant PID2021-122603NB-C21. He further
acknowledges the science research grants from the China Manned Space
Project with NO. CMS-CSST-2021-A01 and CMS-CSST-2021-B01.  We
acknowledge the use of computational resources from the parallel
computing cluster of the Open Physics Hub
(https://site.unibo.it/openphysicshub/en) at the Physics and Astronomy
Department in Bologna.  
TS acknowledges support from FFG grant number 899537.
CJM acknowledges FCT and POCH/FSE (EC) support
through Investigador FCT Contract 2021.01214.CEECIND/CP1658/CT0001.
The figures with the corner plots displaying the posterior
distributions have been produced using the package \texttt{pygtc}
developed by \citet{bocquet16}.

\AckECon
\end{acknowledgements}

\appendix

\bibliography{paper}
\label{lastpage}
\end{document}